\documentclass[aps,10pt,prd,twocolumn,superscriptaddress,preprintnumbers,%
showkeys,nofootinbib,showpacs]{revtex4-2}
\usepackage{amsmath,amsfonts,amssymb,amscd,amsxtra,amsthm}
\usepackage{graphicx}  
\usepackage{epstopdf}
\usepackage{dcolumn}  
\usepackage{bm}          
\usepackage{slashed}
\usepackage{cancel}
\usepackage{float}
\usepackage{mathtools}
\usepackage{amsbsy}
\usepackage{amstext}
\usepackage{hyperref}
\usepackage{wasysym}
\usepackage[utf8]{inputenc} 
\usepackage{booktabs} 
\usepackage[normalem]{ulem} 
\usepackage[dvipsnames]{xcolor} 
\usepackage{tabularx}
\usepackage{enumitem}  
\usepackage{array} 
\usepackage{slashed}
\usepackage{tikz}
\usepackage{float}
\usepackage{multirow}
\usepackage{cancel}
\usepackage{physics}
\makeatletter

\begin{document}  
\title{Tensor-polarized parton density in the $N\to \Delta$ transition from the large-$N_c$ light-cone wave function}  
\author{June-Young Kim}
\email[E-mail: ]{jykim@jlab.org}
\affiliation{Theory Center, Jefferson Lab, Newport News, VA 23606,
  USA} 
\date{\today}
\begin{abstract}
The tensor-polarized parton density is defined by the forward matrix element of a partonic operator in the $N \to \Delta$ transition. In this work, we investigate it by employing the large-$N_c$ light-cone wave function derived from the mean-field approach. The mean-field picture is based on low-energy effective dynamics in the large-$N_c$ limit, where the baryon wave function is formulated in the rest frame.  By exploiting the covariance of the mean-field solution, we derive the corresponding large-$N_c$ light-cone wave function—decomposed unambiguously into $3Q$, $5Q$, $7Q$, and higher Fock components—in the infinite momentum frame. Evaluating the overlap of these wave functions, we derive an overlap representation of the tensor-polarized parton density in the $N \to \Delta$ transition and find that the leading contribution arises from the $5Q$ Fock sector. This indicates that the tensor-polarized parton density directly probes the genuine $5Q$ component and is governed by chiral dynamics. Our numerical analysis shows that the $N \to \Delta$ tensor-polarized parton density is suppressed, consistent with standard large-$N_c$ expectations. Finally, we establish connections among the tensor-polarized parton density, the generalized parton distribution $H_X$, and the energy-momentum tensor form factor $F_4$.
\end{abstract}
\maketitle
\tableofcontents
\section{Introduction}
Generalized parton distributions (GPDs) are a powerful tool for characterizing the nucleon structure in QCD. They parameterize the nucleon matrix elements of non-local QCD operators between the nucleon states. They integrate the concepts of the nucleon parton distribution functions (PDFs) and generalized form factors and provide essential new information about the nucleon structure. For example, the forward limit of the vector GPD is known as the unpolarized PDF, and the first and second $x$-moments of the GPD are related to the vector and energy-momentum tensor~(EMT) form factors, respectively; see Refs.~\cite{Ji:1996nm, Goeke:2001tz, Diehl:2003ny, Belitsky:2005qn, Boffi:2007yc, Lorce:2025aqp} for a review. In particular, the EMT form factors provide information about the spatial distribution of the contributions of quarks/gluons to the mass, spin, and stress in the nucleon~\cite{Polyakov:2002yz, Leader:2013jra, Polyakov:2018zvc, Lorce:2018egm, Burkert:2023wzr, Lorce:2025oot}. 

The definition of nucleon GPDs can be generalized to the $N \to \Delta$ transition, where they parameterize the matrix element of a non-local QCD operator between nucleon and $\Delta$ baryon states. Very recently, a complete set of four independent transition GPDs has been introduced~\cite{Kim:2024hhd}, including the GPD $H_X$, which was omitted in earlier studies~\cite{Goeke:2001tz, Belitsky:2005qn}. From the perspective of the two-dimensional light-front multipole expansion, this new structure corresponds to a quadrupole spin transition $(Q0)$ with no longitudinal angular momentum component ($M=0$). Analogies have been drawn to the tensor-polarized structure of the deuteron; see Refs.~\cite{Kim:2024hhd, Kim:2025ilc} for details. Importantly, this is the only structure that survives in the forward limit ($t, \xi \to 0$) of the $N \to \Delta$ transition matrix element, and the corresponding forward-limit GPD is interpreted as the tensor-polarized parton density~\cite{Kim:2025ilc}. In analogy with the nucleon parton density, this can also be understood as the unpolarized quark distribution, since the same QCD operator probes the unpolarized quark distribution in the nucleon case. Moreover, connections to the transition form factor have been established~\cite{Goeke:2001tz, Belitsky:2005qn, Kim:2023xvw, Kim:2025ilc}, enabling a mechanical interpretation of the transition GPDs. 

Compared to nucleon GPDs, the transition GPDs offer complementary information stemming from the higher-spin nature of the $\Delta$ baryon, thus opening a novel avenue for investigating baryon structure. In particular, only the isovector component of the QCD operator can yield a nonvanishing matrix element, due to the isospin difference between the nucleon and the $\Delta$ baryon, $|I_{\Delta} - I_{N}| = 1$. As a result, the transition GPDs exhibit only a weak dependence on the renormalization scale. This makes them a effective probe of the non-perturbative structure of both the nucleon and the $\Delta$ baryon.

Early studies of the $N \to \Delta$ transition were largely limited to photo- and electroproduction processes, i.e., $N \to \gamma^{(*)} \Delta$. More recently, interest has shifted toward exploring this transition in hard exclusive processes~\cite{CLAS:2023akb, Diehl:2024bmd}. GPDs can be accessed in such processes thanks to QCD factorization, which cleanly separates the non-perturbative content from the perturbative scattering amplitude. They are experimentally measurable through  deeply virtual Compton scattering (DVCS) and deeply virtual meson production (DVMP). Although theoretical and experimental efforts continue to advance~\cite{Diehl:2024bmd, Semenov-Tian-Shansky:2023bsy, Kroll:2022roq, Kroll:2025osx}, much of the dynamical structure of the $N \to \Delta$ transition GPDs remains to be uncovered.

In the study of the $N \to \Delta$ transition matrix element, spin-flavor symmetry in the large-$N_c$ limit of QCD serves as an effective tool, providing a unified framework for describing the matrix elements of QCD operators between the lowest-lying light baryon states—namely, $N \to N$, $N \to \Delta$, and $\Delta \to \Delta$; see Refs.~\cite{tHooft:1973alw,Witten:1979kh,Coleman:1980mx,Gervais:1983wq,Dashen:1993jt}. Large-$N_c$ methods have been extensively applied to analyze $N \to \Delta$ matrix elements of the local vector and axial-vector currents~\cite{Watabe:1995xy, Silva:1999nz, Ledwig:2008es, Ledwig:2008rw, Kim:2020lgp}. They have also been applied to $N \to \Delta$ transition GPDs~\cite{Frankfurt:1999xe, Goeke:2001tz, Pascalutsa:2006up}. 

These standard large-$N_c$ methods can be employed in constructing the large-$N_c$ light-cone wave function, where the dynamical spin-flavor symmetry is adapted to the infinite momentum frame (IMF), and the explicit shape of the wave function depends on the dynamical model. The foundations for the formulation of the large-$N_{c}$ light-cone wave function within the mean-field picture were laid by Petrov and Polyakov in Ref.~\cite{Petrov:2002jr} and elaborated by Diakonov in Refs.~\cite{Diakonov:2004as, Diakonov:2005ib}. Its application to physical observables in terms of the overlap representation has been extensively implemented by Lorcé~\cite{Cedric:2007vc, Lorce:2007xax, Lorce:2007fa, Lorce:2007as, Lorce:2006nq, Lorce:2011dv}. It was also applied to the study of baryon distribution amplitudes~\cite{Kim:2021zbz}. Very recently, this method was applied to the $N \to \Delta$ transition matrix element. The angular momentum content of the partons in the $N \to \Delta$ transition was investigated in Ref.~\cite{Kim:2023yhp}, where the dynamical input for the wave functions was determined by the mean-field picture. In the present work, we aim to extend this analysis to the study of the tensor-polarized parton density in the $N \to \Delta$ transition.

The mean-field picture is based on the low-energy effective dynamics resulting from the spontaneous breaking of chiral symmetry in the QCD vacuum. An explicit realization is provided by the QCD instanton vacuum~\cite{Schafer:1996wv, Diakonov:2002fq}. As a result, the effective theory incorporates the emergence of massless Goldstone bosons (pions) and the dynamical generation of a constituent quark mass. The resulting dynamical quark acquires a mass of approximately $M \approx 300$–$400$~MeV and interacts strongly with the Goldstone pion field, with a dimensionless coupling strength of $M/f_{\pi} \approx 4$.

The low-energy effective theory can be applied to describe baryons in the large-$N_c$ limit of QCD. The underlying idea originates from Witten’s idea~\cite{Witten:1979kh} that a baryon can be viewed as a bound state of $N_c$ valence quarks moving in a classical pion mean field with hedgehog symmetry, since quantum fluctuations of the pion field are parametrically suppressed in the $1/N_c$ expansion. The implementation of this idea within a low-energy effective theory is known as the chiral quark-soliton model, or the mean-field picture.

Importantly, this mean-field picture realizes the dynamical spin-flavor symmetry~\cite{Gervais:1983wq, Dashen:1993jt} in the baryon sector of QCD in the large-$N_c$ limit, where the so-called “hedgehog symmetry” plays a central role in the emergence of this dynamical symmetry. Within this framework, the lowest-lying baryons arise as collective excitations and fall into the octet ($S=I=1/2$) and decuplet ($S=I=3/2$) representations. These states are distinguished by their rotational wave functions. In particular, spin-flavor symmetry provides a unified description of the matrix elements of QCD operators between these light baryon states, including the transitions $N \to N$, $N \to \Delta$, and $\Delta \to \Delta$. 

In this mean-field picture, the baryon wave function at rest is constructed from the $N_c$ valence quark wave function, factorized in color, while the Dirac sea is characterized by a wave function containing an indefinite number of additional quark-antiquark ($Q\bar{Q}$) pairs (see Ref.~\cite{Petrov:2002jr}). As a result, the baryon is described as a superposition of Fock components—$3Q$, $5Q$, $7Q$, and so on. However, the physical baryon wave function at rest simultaneously describes Fock components related to both the baryon and the QCD vacuum, making it impossible to distinguish between them. In contrast, in the IMF, quark–antiquark pairs associated with the baryon carry infinite momentum and are thus clearly separated from those in the vacuum. This distinction allows for a well-defined decomposition of the baryon wave function into $3Q$, $5Q$, $7Q$, and higher Fock components only in the IMF.

To achieve this, the large-$N_c$ wave function must be reformulated in the IMF, which also requires adapting the spin-flavor symmetry to the IMF. This reformulation can be carried out by exploiting the Lorentz invariance of the effective action and the covariance of the mean-field solution~\cite{Petrov:2002jr}. Formulation in the IMF offers an important advantage: a leading-twist QCD operator with finite momentum transfer cannot create or annihilate quarks carrying infinite momentum. As a result, baryon matrix elements are nonzero only between Fock components with equal numbers of quarks and antiquarks. Note, however, that in cases involving finite skewness, transitions between different Fock components become possible.

In this work, we study the forward matrix elements of a non-local vector operator between nucleon states and between nucleon and $\Delta$ baryon states, using the large-$N_c$ light-cone wave function constructed within the mean-field picture. These matrix elements are generically referred to as unpolarized quark distributions when viewed in terms of parton polarization. Alternatively, in the context of baryon polarization, they are referred to as the monopole and tensor-polarized parton densities, respectively.

This paper is organized as follows: In Section~\ref{sec:2}, we parameterize the forward matrix element of a non-local vector operator between the nucleon and $\Delta$ baryon states in terms of unpolarized quark distributions. In Section~\ref{sec:4}, we construct the large-$N_c$ baryon wave function in the rest frame in terms of the $3Q$, $5Q$, $7Q$, ... Fock components within the mean-field picture and derive the corresponding large-$N_c$ light-cone wave function in the IMF, allowing for a precise definition of the baryon’s Fock components. In Section~\ref{sec:5}, using the large-$N_c$ light-cone wave function, we derive overlap representations for the normalization of the baryon wave functions and the unpolarized quark distributions. In Section~\ref{sec:6}, we present and discuss numerical results for the normalization and unpolarized quark distributions. In Section~\ref{sec:7}, we relate the $N \to \Delta$ unpolarized quark distribution to the transition GPD $H_X$~\cite{Kim:2024hhd}, and the transition energy-momentum tensor form factor $F_4$~\cite{Kim:2022bwn}. Lastly, in Section~\ref{sec:8}, we summarize our findings and present our conclusions.

\section{$p\to\Delta^{+}$ transition matrix element of non-local vector operator \label{sec:2}}
In this work, we focus on the specific $p \to \Delta^{+}$ transition as a representative channel of the general $N \to \Delta$ transition. The $p \to \Delta^{+}$ transition GPDs are defined through the matrix element of a non-local vector operator:
\begin{align}
\Gamma_{q} = \bar{\psi}_{q}\left(-\frac{\lambda n}{2}\right) \slashed{n} \psi_{q}\left(\frac{\lambda n}{2}\right),
\label{eq:parton}
\end{align}
with quark flavor $q = u, d, \ldots$, evaluated between proton and $\Delta^{+}$ baryon states:
\begin{align}
\mathcal{M}_{p\Delta^{+}}[\Gamma_{q}] = \int \frac{d\lambda}{2\pi} e^{i\lambda x} \langle \Delta^{+} | \bar{\psi}_{q}\left(-\frac{\lambda n}{2}\right) \slashed{n} \psi_{q}\left(\frac{\lambda n}{2}\right) | p \rangle,
\label{eq:1}
\end{align}
where $n$ is a light-cone vector satisfying $n^2 = 0$, and the light-cone gauge is employed, allowing the gauge link to be omitted. The parameter $\lambda$ denotes the light-cone separation between the non-local quark fields, and $x$ is the momentum fraction carried by the partons. The initial proton state $|p\rangle \equiv |p(p, s)\rangle$ and the final $\Delta^{+}$ baryon state $\langle \Delta^{+}| \equiv \langle \Delta^{+}(p', s')|$ are characterized by momenta $p$ and $p'$, and spin polarizations $s$ and $s'$, respectively. Here, the spin states refer to light-front helicity eigenstates.

The matrix element~\eqref{eq:1} depends on the momentum fraction carried by partons, $x$, the skewness $\xi = - (n \cdot \Delta)/[2(n \cdot P)]$, and the squared momentum transfer $t = \Delta^2$, where the light-cone vector $n$ is normalized by the condition $P \cdot n = 1$. The average and difference of the baryon momenta are defined as
\begin{align}
P= \frac{p'+p}{2}, \quad \Delta= p'-p.
\end{align}
Therefore, Eq.~\eqref{eq:1} can be expressed as a function of the relevant variables:
\begin{align}
\mathcal{M}_{p \Delta^{+}}[\Gamma_{q}] = \text{function}(x,\xi,t;s',s).
\end{align}

Importantly, for a non-diagonal matrix element (e.g., a transition), the forward limit condition $\Delta = 0$ does not apply due to the mass difference between the initial and final states. Instead, the forward limit should be understood as $\xi, t \to 0$; see Ref.~\cite{Kim:2025ilc} and Section~\ref{sec:7} for details. In this forward limit ($\xi, t = 0$), the $p \to \Delta^{+}$ transition matrix element is defined through the tensor-polarized (quadrupole) spin transition:
\begin{align}
\mathcal{M}_{p\Delta^{+}}[\Gamma_{q}] \big{|}_{t,\xi \to 0} \equiv 2 f^{q}_{p\Delta^{+}}(x) (T_{20})_{s's},
\label{eq:4}
\end{align}
where the explicit form of the polarization tensor (see Ref.~\cite{Kim:2025ilc}) is given by
\begin{align}
(T_{20})_{s's} = \left(\begin{array}{r r} 0 & 0 \\[1ex] 1 & 0 \\[1ex] 0 & -1 \\[1ex] 0 & 0  \end{array}\right)_{s's}.
\end{align}
Since the parton density $f_{p\Delta^{+}}(x)$ arises from the tensor-polarized spin structure of the baryon, it is referred to as the tensor-polarized parton density~\cite{Kim:2025ilc}. 

This parton density should be compared to the traditional $p \to p$ parton density. The $p \to p$ transition matrix element of the non-local vector operator~\eqref{eq:parton} is defined by replacing the final $\Delta^{+}$ state in Eq.~\eqref{eq:1} with proton state:
\begin{align}
\mathcal{M}_{pp}[\Gamma_{q}] \big{|}_{\Delta \to 0} \equiv 2 f^{q}_{p}(x) \delta_{s's}.
\label{eq:3}
\end{align}
This expression exhibits the monopole spin structure of the baryon and is therefore referred to as the monopole parton density. Alternatively, both the monopole~\eqref{eq:3} and tensor-polarized~\eqref{eq:4} structures share the same physical interpretation: they describe how much of the baryon's momentum is carried by unpolarized quarks with momentum fraction $x$. Accordingly, both the monopole parton density $f_{p}$ and the tensor-polarized parton density $f_{p\Delta^{+}}$ can be generically referred to as unpolarized quark distributions. In the following, we will refer to them as unpolarized quark distributions.

\section{Mean-field picture \label{sec:4}}
The mean-field picture is based on the low-energy effective dynamics resulting from the spontaneous breaking of chiral symmetry in the QCD vacuum. An explicit realization is provided by the QCD instanton vacuum~\cite{Schafer:1996wv, Diakonov:2002fq}. The resulting low-energy effective chiral Lagrangian is given by
\begin{align}
\mathcal{L}_{\mathrm{eff}} = \bar{\psi}(x) (i\slashed{\partial} - M U^{\gamma_{5}}) \psi(x),
\label{eq:effe}
\end{align}
where $\psi$ and $U^{\gamma_{5}}$ denote the quark and chiral fields, respectively, and $M$ is the dynamical quark mass. The SU(2) chiral field is defined as
\begin{align}
U^{\gamma_{5}} &= \frac{1+\gamma_{5}}{2} U +\frac{1-\gamma_{5}}{2} U^{\dagger},  &&U=\exp[i \bm{\pi}(x)\cdot\bm{\tau}],
\end{align}
where $\bm{\tau}$ denotes the SU(2) flavor matrices. The pion field $\pi(x)$ has hedgehog symmetry, which represents the minimal generalization of spherical symmetry ($r = |\bm{x}|$):
\begin{align}
\pi^{a}(\bm{x}) = n^{a}P(r), \quad n^{a} = x^{a}/r,
\end{align}
and plays a key role in realizing dynamical spin-flavor symmetry in the large-$N_c$ limit. Here, $P(r)$ is a profile function that satisfies the boundary conditions $P(0) = \pi$ and $P(\infty) = 0$.

In the large-$N_c$ limit, one can determine the spectrum of the one-particle Dirac Hamiltonian of the effective theory~\eqref{eq:effe}:
\begin{subequations}
\label{eq:wf_static}
\begin{align}
&H(U) = - i \gamma_{0} \bm{\gamma} \cdot \bm{\nabla}  + \gamma_{0} M U^{\gamma_{5}}, \\[1ex]
&H(U) \Phi_{n} = E_{n} \Phi_{n}.
\end{align}
\end{subequations}
The spectrum consists of a discrete bound-state level and the upper and lower Dirac continua. The nucleon is described as a configuration in which a set of quark single-particle levels (discrete level + lower Dirac continuum) are occupied by $N_c$ quarks, resulting in a state with baryon number $B = 1$. The nucleon mass is obtained by minimizing the energy functional around the saddle point of the pion mean field, which corresponds to the classical nucleon:
\begin{align}
\frac{\delta M_{N}[U]}{\delta U} \bigg{|}_{U=U_{\mathrm{cl}}}= 0.
\label{eq:ns}
\end{align}
The classical solution of the chiral field, $U_{\mathrm{cl}}$, is referred to as the mean-field solution.

Above, we have obtained the standard mean-field solution in the rest frame. As outlined in the Introduction, we now proceed—in the following section—to construct the large-$N_c$ baryon wave function within the mean-field picture. Subsequently, by exploiting the covariance of the mean-field solution~\cite{Petrov:2002jr}, we will briefly outline how the large-$N_c$ light-cone wave function is obtained in the IMF. Throughout this analysis, we closely follow the derivations and notations presented in Refs.~\cite{Petrov:2002jr, Diakonov:2004as, Diakonov:2005ib, Cedric:2007vc, Lorce:2007xax, Lorce:2007fa, Lorce:2007as, Lorce:2006nq, Lorce:2011dv}.

\subsection{Baryon wave function \label{sec:4_1}}
In the large-$N_{c}$ limit, the classical nucleon is expressed as a composition of the $N_{c}$ quark wave functions completely factorized in color. The classical nucleon wave function is then given symbolically as follows (cf.~\cite{Petrov:2002jr, Diakonov:2004as, Diakonov:2005ib}):
\begin{subequations}
\label{eq:LCWF1}
\begin{align}
| \Psi \rangle &= \prod^{N_{c}}_{\mathrm{color}=1} \int (d\bm{p})F(\bm{p})a^{\dagger}(\bm{p}) | \Omega \rangle, \label{eq:LCWF1_a} \\[1ex]
|\Omega \rangle &\equiv  \exp\bigg{[}  \int (d\bm{p}) (d\bm{p}') a^{\dagger}(\bm{p}) W(\bm{p},\bm{p}') b^{\dagger}(\bm{p}')  \bigg{]}| 0 \rangle, \label{eq:LCWF1_b}
\end{align}
\end{subequations}
where $a^{\dagger}$ and $b^{\dagger}$ denote the quark and antiquark creation operators, respectively, and $| \Omega \rangle$ represents the non-trivial QCD vacuum. We will suppress the quantum numbers of the creation operators until they are explicitly needed, and define the momentum integral measure as $(d\bm{p})= d^{3}\bm{p}/(2\pi)^{3}$ for simplicity. The wave function consists of the $3Q$ Fock component in Eq.~\eqref{eq:LCWF1_a} and the higher Fock components $(5Q,7Q,...)$ constructed by expanding the infinite tower of the exponent (quark-antiquark pairs) in Eq.~\eqref{eq:LCWF1_b}. These $3Q$ and vacuum wave functions are characterized by $F(\bm{p})$ and $W(\bm{p},\bm{p}')$, respectively. $F(\bm{p})$ is the bound-state quark wave function~\eqref{eq:wf_static}, and $W(\bm{p},\bm{p}')$ is the quark Green function at equal time in the background of the mean field~\eqref{eq:effe}. These wave functions are determined by the non-trivial solution of the mean-field~\eqref{eq:ns}. Their explicit expressions and concise derivations are given in Sections~\ref{sec:3Q} and \ref{sec:QQbar}.

Importantly, the mean field (or classical nucleon wave function) is degenerate under translations and SU(3) flavor rotations. In order to lift the degeneracy and endow the classical nucleon with momentum and spin-flavor quantum numbers, we consider the translations and rotations of the mean field. After integrating over the translations, we first acquire the momentum sum of the quarks/antiquarks as the baryon momentum. Moreover, the integral over the rotation $R$ leads to the projection of the flavor states of all quarks/antiquarks onto the spin-flavor state $B(R)$, describing the lowest-lying baryon states such as octet and decuplet baryons. Taking these modes into account, we obtain the baryon wave function from Eq.~\eqref{eq:LCWF1} as follows:
\begin{widetext}
\begin{align}
|\Psi^{k}(B) \rangle &=  \int dR \, B^{*}_{k}(R)\frac{\epsilon^{\alpha_{1}\alpha_{2}\alpha_{3}}}{\sqrt{N_{c}!}}  \left[\prod^{N_{c}}_{n=1} \int (d\bm{p}_{n})  R^{f_{n}}_{j_{n}} F^{j_{n}\sigma_{n}}(\bm{p}_{n})a^{\dagger}_{\alpha_{n}f_{n}\sigma_{n}}(\bm{p}_{n})\right]  \cr
& \times  \exp\left( \int (d\bm{p}) (d\bm{p}') a^{\dagger}_{\alpha f \sigma} (\bm{p}) R^{f}_{j} W^{j\sigma}_{j'\sigma'}(\bm{p},\bm{p}') R^{\dagger f'}_{j'} b^{\dagger \alpha f' \sigma'}(\bm{p}')  \right) | 0 \rangle,
\label{eq:LCWF_master}
\end{align}
\end{widetext}
where the quark/antiquark creation operators carry the color $(\alpha=1,2,3)$, spin $(\sigma=1,2=\uparrow, \downarrow)$, and flavor $(f=1,2,3=u,d,s)$ indices. For the spin indices, the creation/annihilation operators are contracted with the quark $F(\bm{p})$ and quark-antiquark pair $W(\bm{p},\bm{p}')$ wave functions. These creation and annihilation operators satisfy the following anticommutation relations:
\begin{align}
&\{a_{\alpha' f' \sigma'}(\bm{p}'), a^{\dagger}_{\alpha f \sigma}(\bm{p}) \} = \{b_{\alpha' f' \sigma'}(\bm{p}'), b^{\dagger}_{\alpha f \sigma}(\bm{p})\} \nonumber \\[1ex]
&= (2\pi)^{3}\delta_{\alpha'\alpha} \delta_{f'f}\delta_{\sigma'\sigma} \delta^{(3)}(\bm{p}' - \bm{p}).
\end{align}
The annihilation operators acting on the vacuum state obviously become zero, i.e., $a,b | 0 \rangle = 0$.

In the first line of Eq.~\eqref{eq:LCWF_master}, the three quarks are antisymmetric in color due to the antisymmetric tensor $\epsilon^{\alpha_{1}\alpha_{2}\alpha_{3}}$. The factor $1/\sqrt{N_{c}!}$ is introduced as a color normalization.
The $3Q$ wave function $F(\bm{p})$ — which accompanies the quark creation operator $a^{\dagger}$ — carries isospin ($j=1,2=u,d$) and spin ($\sigma$) indices. In the isospin indices, the wave functions are contracted with the flavor rotation matrix $R$, which in turn yields the open flavor indices $(f=1,2,3=u,d,s)$, including strangeness. Integrating over this flavor rotation with the rotational wave function $B_{k}(R)$ leads to the baryon states $|\Psi^{k}(B) \rangle$ with spin-flavor quantum numbers. Here, $k$ denotes the spin polarization of the baryon state $B$. The explicit expressions for the spin-flavor rotational wave functions are listed in Appendix~\ref{appendix:a}. Similar to the $3Q$ Fock component, the vacuum wave function is expressed in terms of the quark-antiquark pair wave functions contracted with the flavor rotation matrix $R$. It accompanies the quark $a^{\dagger}$ and antiquark $b^{\dagger}$ creation operators, and importantly, these quark-antiquark pair operators are color-singlet.

\subsection{$3Q$ wave function \label{sec:3Q}}
We now examine how the $3Q$ wave function $F(\bm{p})$ in Eq.~\eqref{eq:LCWF_master} is derived, and construct the light-cone wave function by considering the infinite momentum frame. The leading Fock state of the baryon wave function in Eq.~\eqref{eq:LCWF1} is given by the product of the $N_{c}$ quark creation operators that fill in a bound state~(discrete level):
\begin{subequations}
\label{eq:QWF_0}
\begin{align}
&\prod^{N_{c}}_{\mathrm{color}=1} \int (d^{3} \bm{p}) F(\bm{p}) a^{\dagger}(\bm{p}) | \Omega \rangle  \cr
&\sim \prod^{N_{c}}_{\mathrm{color}=1} \int d^{3}x \, \psi^{\dagger} (x) \Phi_{\rm lev} (\bm{x}) | \Omega \rangle, \label{eq:first} \\[2ex] 
&\text{with} \quad  F(\bm{p}) \equiv F_{\mathrm{lev}}(\bm{p}) + F_{\mathrm{sea}}(\bm{p}). \label{eq:second}
\end{align}
\end{subequations}
There are two different contributions. The quark field operator ${\psi}^{\dagger}$ consists of quark creation and antiquark annihilation operators. The quark creation operator part is directly defined as $F_{\mathrm{lev}}(\bm{p})$, but the antiquark annihilation operator part acts on the quark-antiquark pair in the vacuum state. After considering contractions, the antiquark annihilation operator in ${\psi}^{\dagger}$ is effectively replaced by a quark creation operator through the QCD vacuum, giving rise to the contribution $F_{\mathrm{sea}}(\bm{p})$. This contribution is understood as follows: the one-quark wave function $F_{\mathrm{lev}}(\bm{p})$ is distorted by the amount of $F_{\mathrm{sea}}(\bm{p})$ due to the Dirac sea. In this work, the second term in Eq.~\eqref{eq:second} is ignored because of its complexity. The corrections to the observables have been estimated in Ref.~\cite{Lorce:2011dv} and are found to be about $10\%$.

The bound-state wave function of the quark, $\Phi_{\rm lev} (\bm{x})$, in Eq.~\eqref{eq:QWF_0} is characterized by solving the Dirac equation in the presence of the pion mean field. Note that the bound state of the quark appears only when the mean field is strong enough. The bound state quark wave function has the form
\begin{align}
\Phi_{\rm lev} (\bm{x})
  &= \frac{1}{\sqrt{4\pi}}
\left(
\begin{array}{r} h (r) \\[1ex] 
\displaystyle
-i \frac{\bm{x}\bm{\sigma}}{r} \, j (r) 
\end{array} \right) \chi.
\label{level_spinor}
\end{align}
The bound-state wave function $\Phi_{\rm lev} (\bm{x})$ includes $\chi$, a spinor--isospinor wave function, which satisfies the hedgehog condition $(\bm{\sigma} + \bm{\tau}) \chi = 0$, with the normalization $\chi^\dagger \chi = 1$. The radial components of the wave function, denoted as $h$ and $j$, are governed by the following equation in the chiral limit:
\begin{align}
  &\left(\begin{array}{cc} M \cos{P(r)} & \displaystyle
-\frac{\partial}{\partial r} - \frac{2}{r} + M \sin{P(r)} \\  
           \displaystyle \frac{\partial}{\partial r} + M\sin{P(r)}
& - M\cos{P(r)} \end{array}\right)
  \cr
&\times \left(\begin{array}{c} h(r)  \\[1ex] j(r) \end{array}
  \right)= E_{\mathrm{lev}} \left(\begin{array}{c} h(r)  \\[1ex]
j(r) \end{array}\right).
\label{level_radial}
\end{align} 
The quark wave functions $h(r)$ and $j(r)$ correspond to the orbital angular momentum quantum numbers $l=0$ and $l=1$, respectively. The radial wave functions are normalized as
\begin{align}
\int^{\infty}_{0} dr \, r^{2} \left[h^2(r) + j^2(r) \right] = 1.
\label{level_normalization}
\end{align}
Using the self-consistent mean field $P(r)$, we find that the lower component $j$ accounts for about $20\%$ of the normalization integral~\eqref{level_normalization}, indicating that the relativistic contributions are small but not negligible. $E_{\mathrm{lev}}$ is the quark energy at the discrete level, which is approximately $200$ MeV.

We are now in a position to construct the wave functions $F(\bm{p})$ by making use of Eqs.~\eqref{eq:QWF_0} and \eqref{level_spinor} in the infinite momentum frame. Before discussing this, it is convenient in this frame to rescale the creation and annihilation operators ($a,a^{\dagger},b,b^{\dagger}$) and the integration measure $(d^{3} \bm{p})$ in Eq.~\eqref{eq:QWF_0} [or Eq.~\eqref{eq:LCWF_master}] as follows:
\begin{align}
 \int \frac{d^{3}p}{(2\pi)^{3}} &\to \int dx \int \frac{d^{2}p_{\perp}}{(2\pi)^{2}}, \nonumber \\[1ex]
a^{\dagger}(\bm{p}) &\to a^{\dagger}(x,\bm{p}_{\perp}),
\label{eq:res}
\end{align}
so that the rescaled creation/annihilation operators satisfy the following anticommutation relations:
\begin{align}
&\{a_{\alpha' f' \sigma'}(x',\bm{p}_{\perp}'), a^{\dagger}_{\alpha f \sigma}(x,\bm{p}_{\perp}) \} \nonumber \\[1ex]
&= \{b_{\alpha' f' \sigma'}(x',\bm{p}_{\perp}'), b^{\dagger}_{\alpha f \sigma}(x,\bm{p}_{\perp})\} \nonumber \\[1ex]
&= (2\pi)^{2}\delta_{\alpha'\alpha} \delta_{f'f}\delta_{\sigma'\sigma}\delta(x'-x) \delta^{(2)}(\bm{p}_{\perp}' - \bm{p}_{\perp}).
\label{eq:res_comm}
\end{align}
The momentum fraction carried by a quark is denoted by $x$. Under the above rescaling procedure, Eq.~\eqref{eq:first} becomes
\begin{align}
&\prod^{N_{c}}_{\mathrm{color}=1} \int (d^{3} \bm{p}) F(\bm{p}) a^{\dagger}(\bm{p}) | \Omega \rangle  \cr
&\to \prod^{N_{c}}_{\mathrm{color}=1} \int dx \int \frac{d^{2}p_{\perp}}{(2\pi)^{2}} F(x,\bm{p}_{\perp}) a^{\dagger}(x,\bm{p}_{\perp}) | \Omega \rangle,
\label{eq:QWF_1}
\end{align}
where the remaining factors from the rescaling~\eqref{eq:res} are incorporated into the quark wave function $F(\bm{p})$, which is redefined as $F(x,\bm{p}_{\perp})$.

Considering the infinite momentum frame, taking into account the rescaling, and dropping the distorted Dirac sea contribution to the wave function (i.e., setting $F_{\mathrm{lev}} \equiv F$), the one-quark wave functions in the momentum representation are now derived as
\begin{align}
  &F^{j\sigma}(x,\bm{p}_{\perp}) =  \left(\begin{array}{c c } p_{L}
 f_{\perp}(|\bm{p}|) & f_{\parallel}(|\bm{p}|) \\[2ex] -f_{\parallel}(|\bm{p}|) & p_{R}f_{\perp}(|\bm{p}|) \end{array}\right),
\end{align}
with $p_{R,L}=p_{x}\pm i p_{y}$ and $p_{z}=xM_{N}-E_{\mathrm{lev}}$. Similar to Eq.~\eqref{level_spinor}, there are two independent radial wave functions, $f_{\parallel}$ and $f_{\perp}$. They are expressed in terms of the bound-state wave functions~\eqref{level_radial}:
\begin{align}
f_{\parallel}(\bm{p}) &= \sqrt{\frac{M_{N}}{2\pi}}\left[h(|\bm{p}|) + \frac{p_{z}}{|\bm{p}|}j(|\bm{p}|) \right], \cr
f_{\perp}(\bm{p}) &= \sqrt{\frac{M_{N}}{2\pi}} \left[\frac{j(|\bm{p}|)}{|\bm{p}|} \right],
\label{eq:wf_pp_pl}
\end{align}
where the momentum representations of the bound-state wave functions are obtained via the Fourier transform of Eq.~\eqref{level_spinor}:
\begin{align}
h(|\bm{p}|) &= \int d^{3}x \, e^{-i \bm{p}\cdot \bm{x}} h(r), \cr
\frac{\bm{p}}{|\bm{p}|}j(|\bm{p}|)&= \int d^{3} x \, e^{-i \bm{p}\cdot \bm{x}} \left(-i \frac{\bm{x}}{r}\right) j(r).
\end{align}

\subsection{$Q\bar{Q}$-pair wave function \label{sec:QQbar}}
The quark–antiquark pair wave function $W(\bm{p},\bm{p}')$ is expressed in terms of the finite-time quark Green function at equal times in the presence of the mean field. The external mean field is given by
\begin{align}
&\Pi^{j}_{j'}(\bm{q}) = \int d^{3}\bm{x} \, e^{-i\bm{q}\cdot \bm{x}} (\bm{n}\cdot\bm{\tau})^{j}_{j'} \sin{P(r)}, \cr
&\Sigma^{j}_{j'}(\bm{q}) = \int d^{3}\bm{x} \, e^{-i\bm{q}\cdot \bm{x}} \delta^{j}_{j'} (\cos{P(r)}-1).
\end{align}
The explicit expressions for the pair wave function $W(\bm{p},\bm{p}')$ in the infinite momentum frame are given in Ref.~\cite{Kim:2022ule}. However, they are rather complicated to compute numerically. Therefore, for simplicity, we use the \emph{interpolation approximation}. This can be carried out by expanding the quark–antiquark pair wave function $W(\bm{p}',\bm{p})$ in powers of the gradient of the chiral field $(\partial U \ll M)$. We then obtain an analytic expression in terms of the $U$ field. Typically, the error between the interpolation approximation and the exact calculation lies within at most 15~\%.

To be consistent with the rescaled $3Q$ wave function in Eq.~\eqref{eq:QWF_1} in the infinite momentum frame, we have also rescaled the argument of the vacuum exponent in Eq.~\eqref{eq:LCWF1}. Using Eq.~\eqref{eq:res}, we obtain
\begin{align}
&  \int (d\bm{p}) (d\bm{p}^{\prime}) a^{\dagger}(\bm{p}) W(\bm{p},\bm{p}^{\prime}) b^{\dagger}(\bm{p}^{\prime})  \cr
&\to \int dx  dx'  \frac{d^{2}p_{\perp}}{(2\pi)^{2}} \frac{d^{2}p'_{\perp}}{(2\pi)^{2}} \nonumber \\[1ex]
&\times a^{\dagger}(x,\bm{p_{\perp}}) W(x,\bm{p_{\perp}},x',\bm{p}^{\prime}_{\perp}) b^{\dagger}(x',\bm{p}^{\prime}_{\perp}), 
\end{align}
where the quark–antiquark wave function is rescaled accordingly. In the infinite momentum frame, the rescaled quark–antiquark pair wave function is then derived in Ref.~\cite{Petrov:2002jr} as
\begin{align}
&W^{j \sigma}_{j'\sigma'}(x,\bm{p}_{\perp},x^{\prime},\bm{p}^{\prime}_{\perp})\cr
&= \frac{MM_{N}}{2\pi} \bigg{[} \frac{\Sigma^{j}_{j^{\prime}}(\bm{q})[M(2y-1) + \bm{\mathcal{Q}}_{\perp} \cdot \bm{\tau}]^{\sigma}_{ \sigma^{\prime}}}{ \bm{\mathcal{Q}}^{2}_{\perp} + M^{2}+ y(1-y)\bm{q}^{2}} \cr
&\hspace{1.6cm}+ \frac{i \Pi^{j}_{j^{\prime}}(\bm{q})[M - i\bm{\mathcal{Q}}_{\perp} \times \bm{\tau}_{\perp}]^{\sigma}_{ \sigma^{\prime}}}{ \bm{\mathcal{Q}}^{2}_{\perp} + M^{2}+ y(1-y)\bm{q}^{2}} \bigg{]}.
\end{align}
The longitudinal momentum fractions and transverse momenta for the quark and antiquark are now expressed as the collective 3D momentum of the pair:
\begin{align}
&\bm{q}=(\bm{p}_{\perp}+\bm{p}^{\prime}_{\perp}, (x+x^{\prime})M_{N}),
\end{align}
and the relative transverse momentum between the $Q\bar{Q}$ pair and the longitudinal momentum fraction of the $Q\bar{Q}$ pair carried by the antiquark:
\begin{align}
&\bm{\mathcal{Q}}_{\perp}= \frac{x\bm{p}^{\prime}_{\perp}-x^{\prime}\bm{p}_{\perp}}{x+x^{\prime}}, \quad y= \frac{x^{\prime}}{x+x^{\prime}}.
\end{align}

\section{Overlap representation \label{sec:5}}

\subsection{Normalization \label{sec:5_1}}
We are now in a position to compute the overlap integral of the large-$N_{c}$ light-cone wave functions derived in Sec.~\ref{sec:4}. Before examining the overlap representation of the partonic operator~\eqref{eq:parton} between the baryon states, we first need to normalize the baryon wave functions:
\begin{align}
\langle \Psi^{(n)k} (B) | \Psi(B)^{(n)}_{k} \rangle = \mathcal{N}^{(n)}(B), 
\label{eq:nor}
\end{align}
where $n$ is the number of quarks in the Fock state $(n=3,5,7,...)$ and $k$ is the spin polarization of the baryon. Here, $k$ in Eq.~\eqref{eq:nor} does not denote Einstein summation but refers to the same spin polarization. We will briefly summarize how we determine the normalization of the baryon wave functions.

\begin{figure}[htp]
\centerline{%
\includegraphics[scale=0.175]{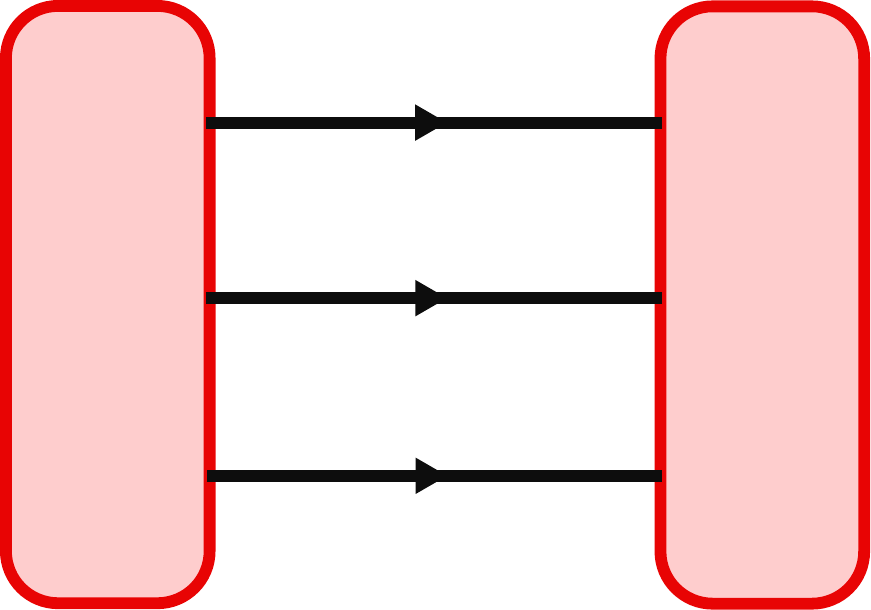} \hspace{0.05cm}
\includegraphics[scale=0.175]{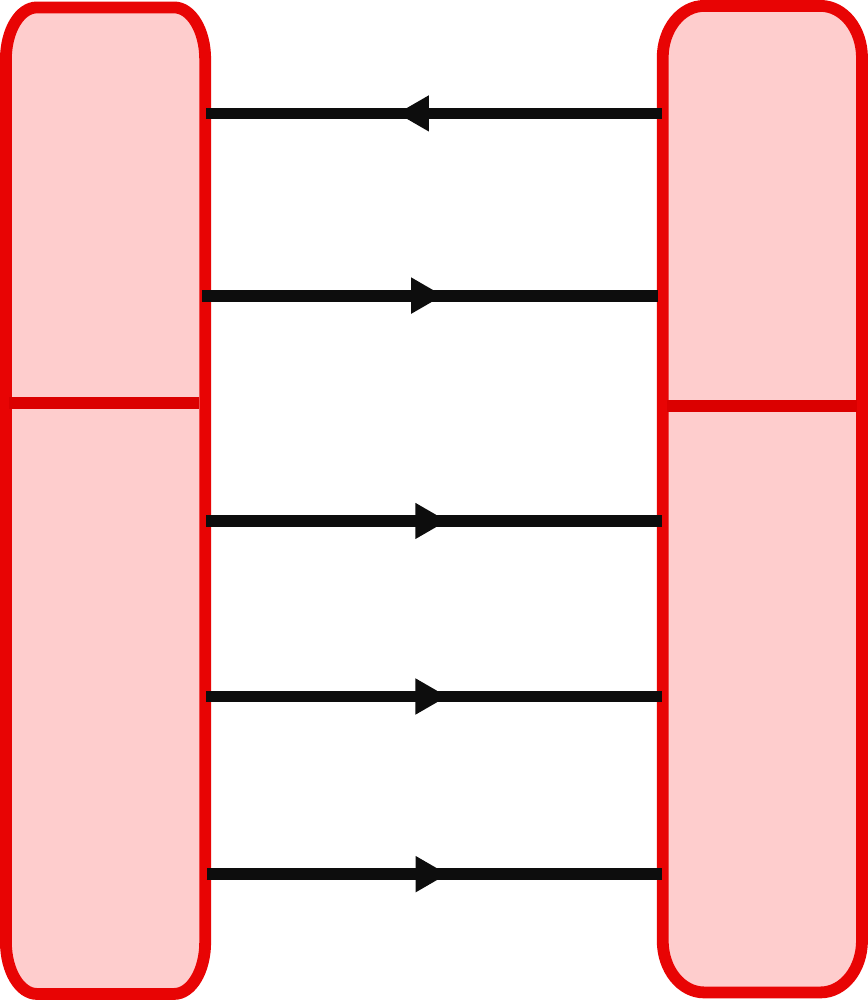} \hspace{0.05cm}
\includegraphics[scale=0.175]{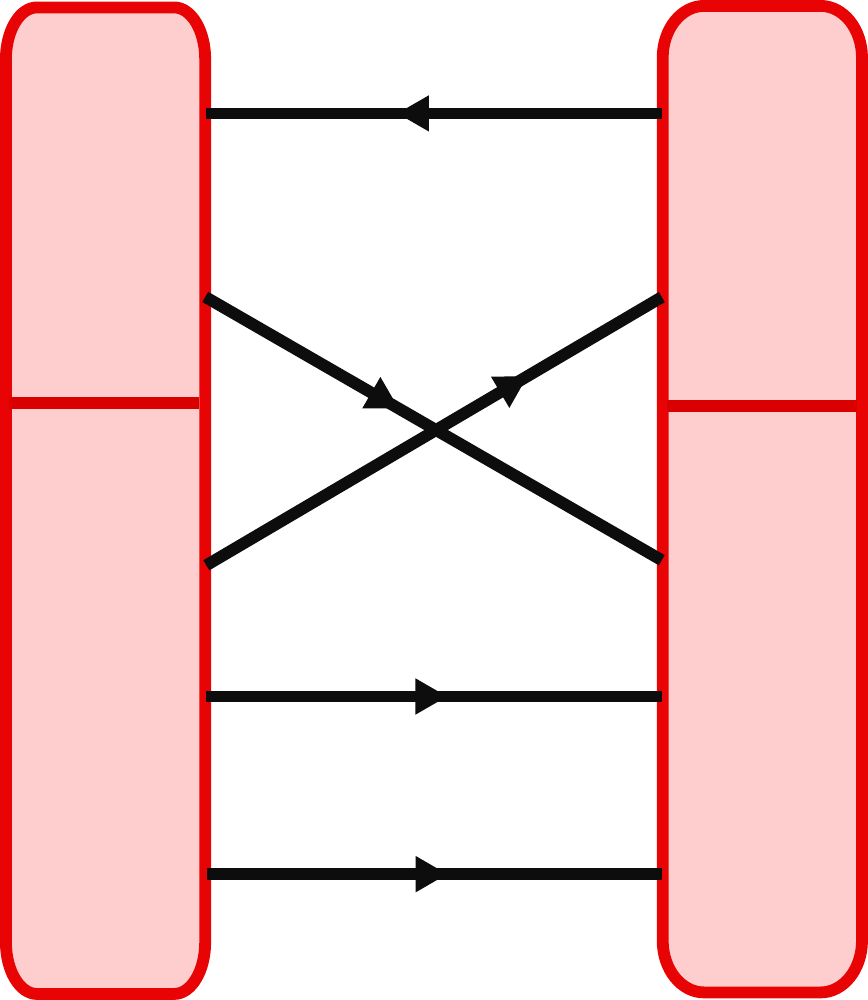}}
\caption{The diagrams of the overlap representations illustrating the normalizations of the $3Q$ (left panel) and $5Q$ (middle and right panels) baryon wave functions are shown. The $5Q$ contributions consist of the ``direct” (middle panel) and ``exchange” (right panel) contributions.}
\label{fig:1_b}
\end{figure}
{\it $3Q$ component.} Having sandwiched the $3Q$ baryon wave functions~\eqref{eq:LCWF_master} and performed the contractions of the creation and annihilation operators, we derive the overlap representation between the $3Q$ baryon wave functions:
\begin{align}
\mathcal{N}^{(3)}(B) &= 6 T(B)^{f_1f_2f_3}_{j_1j_2j_3,k} T(B)_{f_1f_2f_3}^{j'_1j'_2j'_3,k}  \int [d\bm{p}]_{3} \cr
    &\times F^{j_1\sigma_1}(\bm{p}_1)F^{j_2\sigma_2}(\bm{p}_{2})F^{j_3\sigma_3}(\bm{p}_{3}) \cr
    &\times F^{\dagger}_{j'_1\sigma_1}(\bm{p}_1) F^{\dagger}_{j'_2\sigma_2}(\bm{p}_2) F^{\dagger}_{j'_3\sigma_3}(\bm{p}_3), 
\label{eq:3Qnor}
\end{align}
where, from now on, for compactness, we write $F(x,\bm{p}_{\perp})$ as $F(\bm{p})$. The contractions give the global factor of $6$ (left diagram in Fig.~\ref{fig:1_b}). Here, the integration measure for the $nQ$ state is given by
\begin{align}
  \int [d\bm{p}]_n &= \int \left( \prod^{n}_{i=1} dx_i \right) \, \int \left( \prod_{i=1}^n
  \frac{d^{2}\bm{p}_{i\perp}}{(2\pi)^{2}} \right) \cr
  &\times \delta\left(\sum^{n}_{l=1}
  x_{l}-1   \right) 
  (2\pi)^{2}\delta^{(2)}\left(\sum^{n}_{l=1}
  \bm{p}_{l\perp}\right).
\label{eq:Imeasure}
\end{align}
The momentum conservation of the quark/antiquark results from the translational modes of the mean field. Due to the rotational modes of the mean field in Eq.~\eqref{eq:LCWF_master}, each of the three quarks is rotated by the matrix $R^{f}_{j}$. They are then projected onto the spin-flavor baryon state $B^*(R)$ by integration over $R$. The shorthand notation for this group integral is defined as
\begin{align}
  T(B)^{f_{1}f_{2}f_{3}}_{j_{1}j_{2}j_{3},k} \equiv
  \int dR B^{*}_{k}(R)R^{f_{1}}_{j_{1}} R^{f_{2}}_{j_{2}} R^{f_{3}}_{j_{3}}.
\end{align}
We refer to Refs.~\cite{Diakonov:2005ib, Lorce:2006nq, Lorce:2007as,
Lorce:2007xax} for the calculation of the group integral.

To compute $\mathcal{N}^{(3)}(B)$ and to simplify the calculation of the higher Fock state normalization, it is convenient to introduce the probability distribution $\Phi(z,\bm{q}_{\perp})$. It is a function of the longitudinal momentum fraction $z=q_{z}/M_{N}$ and the transverse momentum $\bm{q}_\perp$:
\begin{align}
  &\Phi(z,\bm{q}_{\perp}) = \int \left( \prod^{n}_{i=1} dx_i \right)\, \delta\left(\sum^{3}_{l=1}
                           x_{l}+z-1\right) \cr
                           &\times \int \left(\prod_{i=1}^3
                           \frac{d^{2}\bm{p}_{i\perp}}{(2\pi)^{2}}\right)   (2\pi)^{2}\delta^{(2)}\left(\sum^{3}_{l=1}\bm{p}_{l\perp} +\bm{q}_{\perp}\right) \nonumber \\[1ex]
                           &\times D(\bm{p}_1,\bm{p}_2,\bm{p}_3),
\label{eq:valence_prob}
\end{align}
where the function $D(\bm{p}_1,\bm{p}_2,\bm{p}_3)$ is defined as
\begin{align}
&D(\bm{p}_1,\bm{p}_2,\bm{p}_3) \equiv \bigg{[}f^{2}_{\parallel}(\bm{p}_1)+\bm{p}^{2}_{2 \perp}f^{2}_{\perp}(\bm{p}_1)\bigg{]}\cr
&\times\bigg{[}f^{2}_{\parallel}(\bm{p}_2)+\bm{p}^{2}_{2 \perp} f^{2}_{\perp}(\bm{p}_2)\bigg{]}\bigg{[}f^{2}_{\parallel}(\bm{p}_3)+\bm{p}^{2}_{3 \perp}f^{2}_{\perp}(\bm{p}_3)\bigg{]}.
\label{eq:valence_prob_func}
\end{align}
The probability distribution~\eqref{eq:valence_prob} gives information about how the $3Q$ leave the longitudinal momentum $z=q_{z}/M_{N}$ and the transverse momentum $\bm{q}_{\perp}$. When considering the $3Q$ Fock states, these momentum leakages become zero $(\bm{q}=0)$ due to the conservation of momentum. Thus, the probability distribution at $\bm{q}=0$, i.e., $\Phi(0,0)$, parameterizes the $3Q$ normalization~\eqref{eq:3Qnor}. After summing over all indices in Eq.~\eqref{eq:3Qnor}, we get the $3Q$ normalizations in terms of the probability distribution $\Phi(0,0)$:
\begin{align}
&\mathcal{N}^{(3)}(B_{8})=\frac{3}{2}\Phi(0,0), \cr
&\mathcal{N}^{(3)}_{1/2}(B_{10})=\mathcal{N}^{(3)}_{3/2}(B_{10})=\frac{3}{5}\Phi(0,0),
\label{normalization_3Q}
\end{align}
where $B_{8}$ and $B_{10}$ denote the octet and decuplet baryons, and $1/2$ and $3/2$ in the subscripts of $\mathcal{N}^{(3)}$ refer to the spin polarization of the decuplet baryons. Note that the normalization of the discrete-level wave functions $f_{\perp}$ and $f_{\parallel}$ is arbitrary. We choose it such that $\Phi(0,0)=1$.

When considering the higher-Fock states, this outflow of momentum $\bm{q}$ is taken up by the 3D collective momentum of the quark-antiquark pair, and the total quark/antiquark momentum is conserved again. Thus, the $5Q$ wave function can be viewed as follows: the information in the quark-antiquark pair wave function is almost factorized from the probability distribution $\Phi$ representing the $3Q$ wave function, and they communicate the internal dynamics through the variable $\bm{q}$. Thus, one can see that introducing the probability distribution is useful in computing the normalization for the $5Q$ Fock component and facilitates the physical interpretation.

{\it $5Q$ component.} After sandwiching the $5Q$ baryon wave functions~\eqref{eq:LCWF_master} and contracting the creation and annihilation operators, we derive the overlap representation between the $5Q$ baryon wave functions:
\begin{widetext}
\begin{align}
\mathcal{N}^{(5)}(B) &= 18 T(B)^{f_{1}f_{2}f_{3}f_4,j_5}_{j_{1}j_{2}j_{3}j_4,f_5,k} T(B)^{j'_{1}j'_{2}j'_{3}j'_4,f_5,k}_{f_{1}f_{2}g_{3}g_4,j'_5} \int [d\bm{p}]_5  \,  F^{j_1\sigma_1}(\bm{p}_1)F^{j_2\sigma_2}(\bm{p}_2)F^{j_3\sigma_3}(\bm{p}_3) \nonumber \\[1ex]
&\times W^{j_4 \sigma_4}_{j_5\sigma_5}(\bm{p}_{4},\bm{p}_{5})  F^{\dagger}_{j'_1\sigma_1}(\bm{p}_1) F^{\dagger}_{j'_2\sigma_2}(\bm{p}_2)  \bigg{[}F^{\dagger}_{j'_3\sigma_3}(\bm{p}_3)W^{ j'_5 \sigma_{5} \dagger}_{j'_4\sigma_4}(\bm{p}_{4},\bm{p}_{5}) \delta^{g_{3}}_{f_{3}}\delta^{g_{4}}_{f_{4}} - F^{\dagger}_{j'_3\sigma_4}(\bm{p}_4)W^{ j'_5 \sigma_{5} \dagger}_{j'_4\sigma_3}(\bm{p}_{3},\bm{p}_{5}) \delta^{g_{3}}_{f_{4}}\delta^{g_{4}}_{f_{3}}\bigg{]},
\label{eq:5Qnor}
\end{align}
\end{widetext}
where we rewrite $W(x,\bm{p}_{\perp},x^{\prime},\bm{p}^{\prime}_{\perp})$ as $W(\bm{p},\bm{p}^{\prime})$ for the sake of compactness. There are two contributions. One is the contribution from the ``direct" quark exchange~(middle diagram in Fig.~\ref{fig:1_b}) between the initial and final wave functions. While the direct exchange of the initial and final quarks in $3Q$ configurations gives a combinatorial factor of $6$, the direct quark exchange between the color-singlet quark-antiquark pairs gives an additional combinatorial factor $N_{c}=3$. In total, the ``direct" contribution has a combinatorial factor of $6 \cdot 3$, which is represented by the first term in Eq.~\eqref{eq:5Qnor}.

Another contribution arises from the cases where a quark is ``exchanged" between the $3Q$ and the $Q\bar{Q}$ pair~(right diagram in Fig.~\ref{fig:1_b}). One of the $3Q$ quarks is exchanged with the quark in the $Q\bar{Q}$ pair, giving a combinatorial factor $3 \cdot 3$. In addition, the number of cases for quark exchange among the remaining quarks in the $3Q$ configuration gives an additional factor of $2$. In total, the ``exchange" contribution has a combinatorial factor of $6 \cdot 3$, which is the same as that of the ``direct" contribution. It is represented by the second term in Eq.~\eqref{eq:5Qnor}.

Note that this term has the opposite sign to the ``direct" contribution, resulting from the exchange of creation/annihilation operators. However, in this work, we will only consider the ``direct" contributions. This assumption can be justified because the exchange diagram is parametrically suppressed in the large-$N_{c}$ limit. Numerically, it has been shown that the exchange contribution is about 1–2\% of the direct contribution, and that this contribution is indeed negligible.

\begin{figure}[htp]
\centerline{%
\includegraphics[scale=0.25]{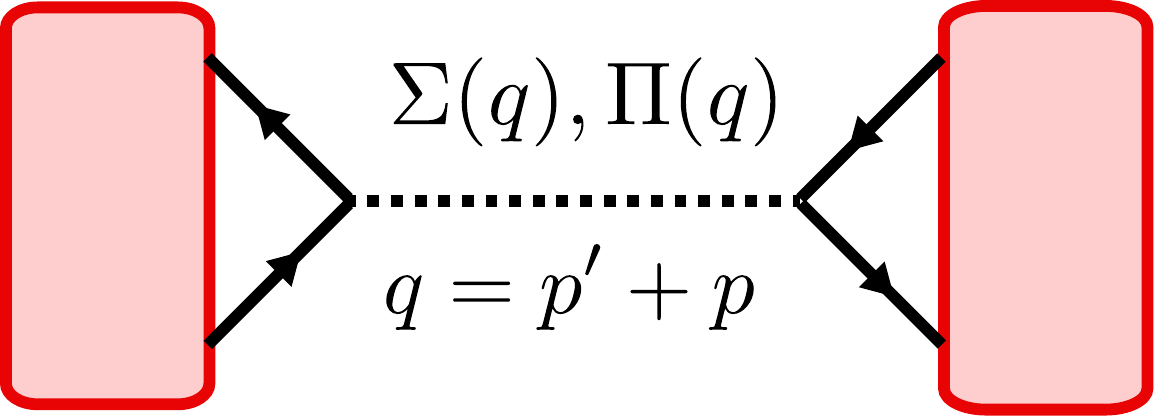}}
\caption{The portion of the quark-antiquark pair contributions to the $5Q$ normalization~\eqref{eq:5Qnor} in the overlap representation is shown diagrammatically and is related to the self-consistent mean fields $\Sigma$ and $\Pi$. The momenta $\bm{q}$ of the mean fields $\Sigma$ and $\Pi$ are given by the sum of the quark momentum $\bm{p}$ and antiquark momentum $\bm{p}'$.}
\label{fig:1_q}
\end{figure}
In the case of the direct contributions, the sum of the pair wave functions in spin $(\sigma_{4},\sigma_{5})$ is given by
\begin{align}
W^{j_4 \sigma_4}_{j_5 \sigma_5} W^{j^{\prime}_5 \sigma_5 \dagger}_{j^{\prime}_4 \sigma_4} &= \delta^{j_{4}}_{j_{5}}  \delta^{j^{\prime}_{5}}_{j^{\prime}_{4}} \mathcal{G}_{\sigma\sigma} (\bm{q},y,\bm{Q}_{\perp}), \cr
& + (\tau^{a})^{j_4}_{j_5}(\tau^{b})^{j^{\prime}_5}_{j^{\prime}_4} \frac{q^{a}q^{b}}{|\bm{q}|^{2}} \mathcal{G}_{\pi\pi}(\bm{q},y,\bm{Q}_{\perp}),
\label{eq:sumpair}
\end{align} 
where the quark-antiquark pair contributions are interpreted as two distinct $\Sigma$ and $\Pi$ exchange contributions~(see Fig.~\ref{fig:1_q}). They are defined as
\begin{align}
\mathcal{G}_{\sigma\sigma} (\bm{q},y,\bm{\mathcal{Q}}_{\perp}) &=  \frac{2\Sigma^{2}(\bm{q})\left[\bm{\mathcal{Q}}^{2}_{\perp} + M^{2}(2y-1)^{2}\right]}{(\bm{\mathcal{Q}}^{2}_{\perp} + M^{2} + y(1-y) \bm{q}^{2})^{2}},  \cr
\mathcal{G}_{\pi\pi} (\bm{q},y,\bm{\mathcal{Q}}_{\perp}) &=  \frac{2\Pi^{2}(\bm{q})\left[\bm{\mathcal{Q}}^{2}_{\perp} + M^{2}\right]}{(\bm{\mathcal{Q}}^{2}_{\perp} + M^{2} + y(1-y) \bm{q}^{2})^{2}}. 
\label{eq:density}
\end{align}
Combining Eqs.~\eqref{eq:sumpair} and \eqref{eq:valence_prob}, we express the $5Q$ normalization of the wave function as the dynamical parameter $K_J$. It is defined by the integral of the factorized form of the probability distribution~\eqref{eq:valence_prob} and $Q\bar{Q}$ pair wave functions $G_{J}$ over $\bm{q}$:
\begin{align}
K_{J} = \frac{M^{2}}{2\pi} \int \frac{d^{3}q}{(2\pi)^{3}} \Phi(\frac{q_{z}}{M_{N}}, \bm{q}_{\perp}) \theta(q_{z}) q_{z} G_{J}(\bm{q}),
\label{eq:nor_para}
\end{align}
where the $G_{J}(\bm{q})$ ($J=\sigma\sigma,\pi\pi,33$) denote the meson exchange contributions as a function of the 3D momentum of the meson~($Q\bar{Q}$-pair). They are defined by Eq.~\eqref{eq:density}, with the longitudinal momentum fraction of the antiquark $y$ and the relative transverse momentum $\bm{\mathcal{Q}}_{\perp}$ integrated out:
\begin{subequations}
\label{eq:dy}
\begin{align}
G_{J}(\bm{q}) &=\int^{1}_{0} dy \int \frac{d^{2}\bm{\mathcal{Q}}}{(2\pi)^{2}} \bigg{[} \mathcal{G}_{J}(\bm{q},y,\bm{\mathcal{Q}}_{\perp})  \bigg{]} \bigg{|}_{\mathrm{reg}}, \label{eq:dy1} \\[1ex]
&\text{with} \quad \mathcal{G}_{33} \equiv \frac{q^{2}_z}{|\bm{q}|^{2}} \mathcal{G}_{\pi \pi}.
 \label{eq:dy2}
\end{align}
\end{subequations}
Since $G_J$ are logarithmically divergent, we introduce the Pauli-Villars regulator. The regularization is implemented as follows:
\begin{align}
f(M) \bigg{|}_{\mathrm{reg}} \equiv f(M) - f(M_{\mathrm{PV}}).
\label{eq:reg}
\end{align}
Since the 3D rotational symmetry is broken in the overlap representation~\eqref{eq:5Qnor}, the additional integral of the $q^{2}_{z}$-weighted $\mathcal{G}_{\pi \pi}$ distribution~\eqref{eq:sumpair} appear as an independent dynamical parameter $G_{33}$~\eqref{eq:dy2}.

After summing over all the spin/isospin/flavor indices in the overlap representation~\eqref{eq:5Qnor} and projecting them onto the spin-flavor baryon states $B(R)$, one arrives at the $5Q$ normalization constant $\mathcal{N}^{(5)}$ expressed in terms of $K_J$ ($J=\sigma\sigma,\pi\pi,33$):
\begin{align}
\mathcal{N}^{(5)}(B_{8}) &= \frac{3}{10}(11K_{\pi \pi} + 23K_{\sigma \sigma}), \cr
  \mathcal{N}^{(5)}_{1/2}(B_{10}) &= \frac{3}{20}(11K_{\pi \pi} + 6 K_{33} +
    17K_{\sigma \sigma}), \cr
    \mathcal{N}^{(5)}_{3/2}(B_{10})
    &= \frac{3}{20}(15K_{\pi \pi} - 6 K_{33} + 17K_{\sigma \sigma}).
\end{align}

\subsection{Quark distribution function \label{sec:5_2}}
Using the overlap representation of the large-$N_{c}$ light-cone wave functions, we now compute the quark distribution. The great advantage of the infinite momentum frame is that the number of $Q\bar{Q}$ pairs is not changed by the current, so only diagonal transitions in the Fock space occur. In this context, the quark distribution can be decomposed into contributions from individual Fock components as follows:
\begin{align}
f^{q}_{B B'}(x) = \sum^{\infty}_{n=3,5,...} f^{q, (n)}_{B B'}(x),
\end{align}
where the contributions of the individual Fock components are defined by the properly normalized overlap representation $O^{(n)}$:
\begin{align}
&f^{q, (n)}_{B B'}(x) = \frac{O^{q, (n)}_{B B'}(x) }{\sqrt{\sum\limits^{n}_{m=3,5,...}\mathcal{N}^{(m)}(B)}\sqrt{\sum\limits^{n}_{l=3,5,...}\mathcal{N}^{(l)}(B')}}.
\label{normalization}
\end{align}
In practice, one must truncate the Fock expansion at the component level where the observables begin to saturate. In this work, we consider the $3Q$ (leading) and $5Q$ (subleading) Fock components in estimating the quark distribution.

Before discussing the results for the quark distributions, we need to clarify the different definitions of the valence and sea quark contributions to the quark distributions in the present framework compared to QCD. First, depending on whether the QCD operator acts on the $3Q$ component or the $Q\bar{Q}$ pair, the quark distributions are split into valence and sea quark contributions. They are defined as
\begin{align}
&f^{q}(x) = f^{q_{\mathrm{val}}}(x) + f^{q_{\mathrm{sea}}}(x), \nonumber \\[1.5ex]
&\text{with} \quad f^{q_{\mathrm{sea}}}(x)= f^{q_\mathrm{s}}(x) + f^{\bar{q}}(x),
\label{eq:decom_vs}
\end{align}
where $f^{q_{\mathrm{val}}}$ is the valence quark contribution (active $3Q$), and $f^{q_{\mathrm{sea}}}$ denotes the sea quark contribution (active $Q\bar{Q}$ pair). The sea quark contribution can be further split into the quark $f^{q_\mathrm{s}}$ and antiquark $f^{\bar{q}}$ contributions of the $Q\bar{Q}$ pair. Note that the supports of the quark ($f^{q}$, $f^{q_\mathrm{s}}$) and antiquark ($f^{\bar{q}}$) distributions are given by $x = [0,1]$ and $x = [-1,0]$, respectively.

On the other hand, the conventional QCD valence quark $f^{q_{\mathrm{v}}}$ is defined by adding the antiquark contributions $f^{\bar{q}}$ to the total quark contributions $f^{q_\mathrm{tot}}$:
\begin{align}
&f^{q_{\mathrm{v}}}\equiv f^{q_\mathrm{tot}}+f^{\bar{q}} = f^{q}, \nonumber \\[1.5ex]
&\text{with} \quad f^{q_\mathrm{tot}}\equiv f^{q_\mathrm{val}} + f^{q_\mathrm{s}}.
\end{align} 
In this definition, the sea quark pairs are commonly thought to be produced in the perturbative process of gluon~(flavor singlet) splitting, leading to
\begin{align}
f^{q_{\mathrm{s}}}(x)=-f^{\bar{q}}(-x), \quad \text{with} \quad 0<x<1.
\label{eq:as}
\end{align}
Thus, the net number of quarks $N^{q}$ inside a baryon is determined solely by the QCD valence quarks:
\begin{align}
&\int^{1}_{-1} dx \, f^{q}(x)= \int^{1}_{-1} dx \, f^{q_{\mathrm{v}}}(x) = N^{q}, \nonumber \\[0.5ex]
&\text{with} \quad \int^{1}_{-1} dx \, f^{q_{\mathrm{sea}}}(x) = 0.
\label{eq:QCD_de}
\end{align}
According to the assumption of Eq.~\eqref{eq:as}, one can easily see that the definition of the valence quark contribution in the present framework is equivalent to that in QCD, leading to $f^{q_\mathrm{v}}(x)= f^{q_{\mathrm{val}}}(x)$. However, the perturbative assumption~\eqref{eq:as} does not always hold in non-perturbative dynamics. When we consider the $5Q$ component of the proton wave function in the present framework, non-standard $3Q$ configurations — which differ from the standard valence quark configuration $| uud (q\bar{q}) \rangle$ — are allowed, such as $| udd (u\bar{d}) \rangle$. This is possible in non-perturbative QCD due to the creation of the pion (a flavor non-singlet) in the QCD vacuum. Thus, the flavor asymmetry results in $f^{q_{\mathrm{s}}}(x) \neq -f^{\bar{q}}(-x)$, leading to $f^{q_\mathrm{v}}(x)\neq f^{q_{\mathrm{val}}}(x)$. Consequently, one should keep in mind that we follow the definitions of the valence and sea quarks given in Eq.~\eqref{eq:decom_vs} throughout this work.

\begin{figure}[htp]
\centerline{%
\includegraphics[scale=0.20]{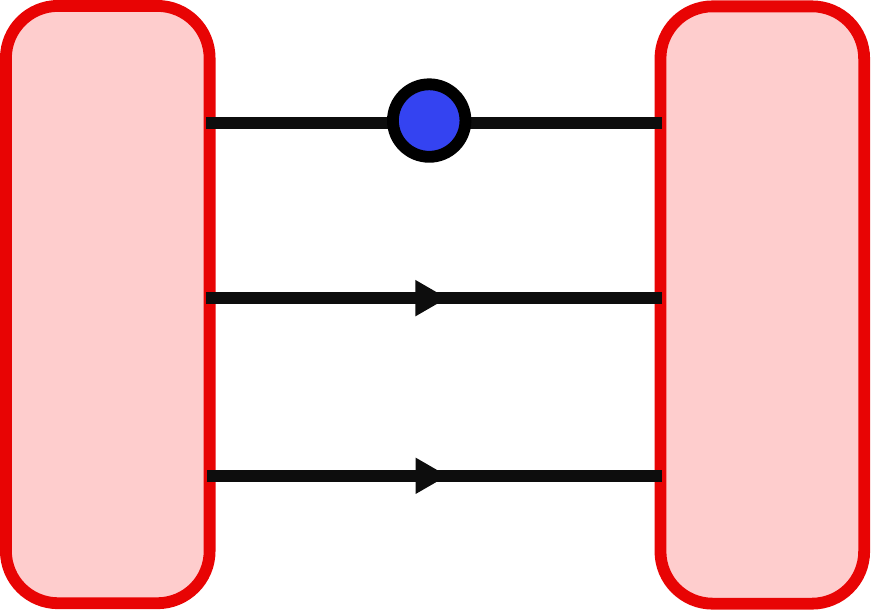}}
\caption{The overlap representation between the $3Q$ wave functions for the quark distribution is drawn diagrammatically. The quark lines are indicated by black lines with arrows, and the non-local operator acting on these quarks is denoted by a blue circle.}
\label{fig:1_c}
\end{figure}

{\it $3Q$ component.} We first derive the overlap representation for the quark distributions in the leading Fock components:
\begin{align}
&O^{q,(3)}_{B B'}(x) =  18   T(B')_{qf_2f_3}^{j'_{1}j'_{2}j'_{3},k} T(B)^{qf_2f_3}_{j_{1}j_{2}j_{3},k}   \int[d\bm{p}]_{3} \cr
&\times \delta(x-x_{1})    F^{j_1\sigma_1}(\bm{p}_1)F^{j_2\sigma_2}(\bm{p}_2)F^{j_3\sigma_3}(\bm{p}_3) \nonumber \\[1ex]
&\times F^{\dagger}_{j'_1\sigma_1}(\bm{p}_1)F^{\dagger}_{j'_2\sigma_2}(\bm{p}_2)F^{\dagger}_{j'_3\sigma_3}(\bm{p}_3). 
\label{eq:UPDFs3Q}
\end{align}
The QCD quark operator $a^{\dagger} a$ is applied to the quark from the $3Q$ wave function, giving a combinatorial factor of $3\cdot 3$. The number of permutations  among the remaining quarks gives a factor of $2$. In total, the contractions yield $3\cdot 3 \cdot 2$. The corresponding schematic diagram is shown in Fig.~\ref{fig:1_c}.

Similar to the normalization, it is convenient to introduce the probability distribution. In particular, in order to adapt it for the quark distribution, we generalize it to a function of the partonic variable $x$, i.e., $\phi(x, z,\bm{q}_{\perp})$. Thus, the generalized probability distribution is given as a function not only of the longitudinal momentum fraction $z=q_{z}/M_{N}$ and of the transverse momentum $\bm{q}_\perp$, but also of the partonic variable $x$:
\begin{align}
  &\phi(x,z,\bm{q}_{\perp}) =\int \left( \prod^{n}_{i=1} dx_i \right)  \delta\left(\sum^{3}_{l=1}
                           z_{l}+z-1\right)   \cr
                           &\times  \int \left(\prod_{l=1}^3
                           \frac{d^{2}\bm{p}_{l\perp}}{(2\pi)^{2}}\right) (2\pi)^{2}\delta^{(2)}\left(\sum^{3}_{l=1}\bm{p}_{l\perp} +\bm{q}_{\perp}\right) \nonumber \\[1ex]
                           &\times \delta(x-x_{1})  D(\bm{p}_1,\bm{p}_2,\bm{p}_3) ,
\label{eq:valence_prob_2}
\end{align}
where $D$ is defined in Eq.~\eqref{eq:valence_prob_func}. Importantly, the first $x$-moment of the generalized probability distribution~\eqref{eq:valence_prob_2} is normalized to the probability distribution~\eqref{eq:valence_prob}:
\begin{align}
\int dx \, \phi(x,z,\bm{q}_{\perp}) = \Phi(z,\bm{q}_{\perp}).
\label{eq:nor_pdf}
\end{align}
Like the probability distribution, the generalized probability distribution gives information about how the $3Q$ leave the longitudinal momentum $z$ and the transverse momentum $\bm{q}_{\perp}$. The only difference is the appearance of the Dirac delta function in the integrand~\eqref{eq:valence_prob_2}, indicating that one of the $3Q$ is activated by an external current. When considering the $3Q$ Fock states, these momentum leakages become zero $(\bm{q} = 0)$ due to conservation of momentum. Thus, at the leading Fock component, the proton and $\Delta^{+}$ baryon quark distributions are determined only by the generalized probability distribution $\phi(x,q_{z}/M_{N},\bm{q}_{\perp})$ at $\bm{q}=0$. After summing over the spin/flavor/isospin indices in Eq.~\eqref{eq:UPDFs3Q}, we derive the separate flavor quark distributions for the proton:
\begin{align}
&O^{u_{\mathrm{val}},(3)}_{p}(x) = 2 O^{d_{\mathrm{val}},(3)}_{p}(x) = 3 \phi(x,0,0),  \cr
&O^{s_{\mathrm{val}},(3)}_{p}(x)= 0,
\label{eq:over_3Q_p}
\end{align}
and those for the $\Delta^{+}$ baryon
\begin{align}
&O^{u_{\mathrm{val}},(3)}_{\Delta^{+}} (x)= 2 O^{d_{\mathrm{val}},(3)}_{\Delta^{+}}(x) = \frac{6}{5}\phi(x,0,0), \cr
&O^{s_{\mathrm{val}},(3)}_{\Delta^{+}}(x) = 0.
\label{eq:over_3Q_d}
\end{align}
Indeed, we observe that these quark distributions are expressed only by $\phi(x,0,0)$. The valence $u$-quark contribution is twice as large as the $d$-quark contribution, reflecting the net number of $u$ and $d$ quarks inside the proton and the $\Delta^{+}$ baryon, i.e., $| u u d \rangle$. Since there are no strange quarks at the valence level, the strange quark contribution is zero for both the proton and the $\Delta$ baryon. Obviously, there are no sea quark or antiquark contributions:
\begin{align}
O^{q_{s},(3)}_{p,\Delta^{+}}=0, \quad O^{\bar{q},(3)}_{p,\Delta^{+}}=0,
\end{align}
due to the absence of the $Q\bar{Q}$ pair.

The concept of the quark distributions for the nucleon and the $\Delta$ baryon can be extended to the transition quark distribution by replacing the diagonal ($N \to N$, $\Delta \to \Delta$) rotational wave functions with the non-diagonal ($N \to \Delta$) ones. Remarkably, we find that the quark distribution in the $N \to \Delta$ transition becomes zero in the overlap representation of the $3Q$ Fock component:
\begin{align}
O^{q,(3)}_{p\Delta^{+}}(x) &= 0.
\end{align}
This is due to the dynamical spin-flavor symmetry adapted to the infinite momentum frame.

Next, by normalizing the results of the overlap representations with the baryon wave function normalizations~\eqref{normalization_3Q}, we derive the physical quark distributions~\eqref{normalization}. By truncating the Fock expansion to the leading component, we define the physical quark distribution as
\begin{align}
&f^{q}_{B B'} = \frac{O^{q, (3)}_{B B'}}{\sqrt{\mathcal{N}^{(3)}(B)} \sqrt{\mathcal{N}^{(3)}(B')}}.
\label{eq:3Qnor_phy}
\end{align}

By substituting the results of the overlap representations in Eqs.~\eqref{eq:over_3Q_p} and \eqref{eq:over_3Q_d} and the baryon wave function normalizations~\eqref{normalization_3Q} into Eq.~\eqref{eq:3Qnor_phy}, we derive the physical quark distributions for the proton and the $\Delta^{+}$ baryon:
\begin{align}
&\left\{ \begin{array}{c} f^{u}_{p}(x) \\[2ex] f^{u}_{\Delta^{+}}(x) \end{array} \right\} = \left\{ \begin{array}{c} 2 f^{d}_{p}(x) \\[2ex]
2 f^{d}_{\Delta^{+}}(x) \end{array} \right\}   = 2\frac{\phi(x,0,0)}{\Phi(0,0)}.
\end{align}
Interestingly, the physical quark distributions for the proton and the $\Delta^{+}$ baryon are found to be identical. Importantly, since the first $x$-moment of the non-local vector current becomes the local vector current, that of the quark distribution should also correspond to the vector charge, which is a conserved quantity. In the present framework, using the relation~\eqref{eq:nor_pdf}, we are able to confirm that the first $x$-moment of the quark distributions reproduces the vector charges (i.e., the net number of valence quarks) as follows:
\begin{align}
\frac{1}{2}\left\{ \begin{array}{c} \int^{1}_{-1} dx \, f^{u}_{p}(x) \\[2ex]  \int^{1}_{-1} dx \, f^{u}_{\Delta^{+}}(x) \end{array} \right\} = \left\{ \begin{array}{c} \int^{1}_{-1} dx \, f^{d}_{p}(x) \\[2ex]
\int^{1}_{-1} dx \, f^{d}_{\Delta^{+}}(x) \end{array} \right\} 
 = 1.
\end{align}

Moreover, the second $x$-moment of the non-local vector current is related to the energy-momentum tensor operator, i.e., a conserved quantity for flavor singlet. Therefore, the second moment of the quark distribution is strongly constrained by global symmetry and normalized to unity, i.e., $x_{1}+x_{2}+x_{3}=1$ (momentum sum rule).
It is straightforward to verify that the integrals of the $x$-weighted quark distributions for the proton and the $\Delta^{+}$ baryon over $x$ give the correct momentum sum rules:
\begin{align}
\int^{1}_{-1} dx  \, x f^{u + d + s}_{p,\Delta^{+}}(x) = 1.
\end{align}
On the other hand, there is no known global symmetry in the flavor non-singlet components, so there is no known sum rule. This implies that they are fully determined by the underlying dynamics of the model.

Unlike the results for the diagonal matrix elements of the non-local operator, we obtain a null result for the $p\to \Delta^{+}$ transition:
\begin{align}
&f^{q}_{p\Delta^{+}}(x) = 0.
\end{align}
Provided that the perturbative treatment of the Fock expansion works, we can conclude that the quark distribution in the $p\to \Delta^{+}$ transition is very small compared to that in the proton. This observation highlights the need to look beyond the leading Fock component when studying the quark distribution in the $p\to \Delta^{+}$ transition.

\begin{figure}[htp]
\centerline{%
\includegraphics[scale=0.175]{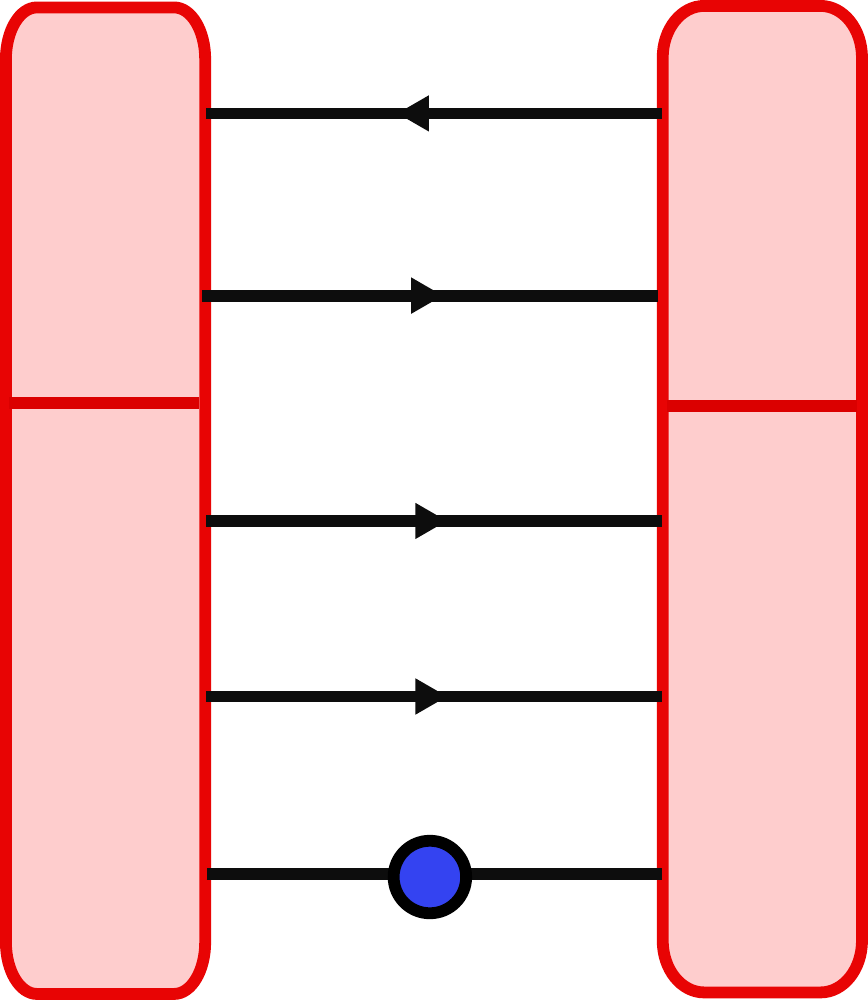} \hspace{0.1cm}
\includegraphics[scale=0.175]{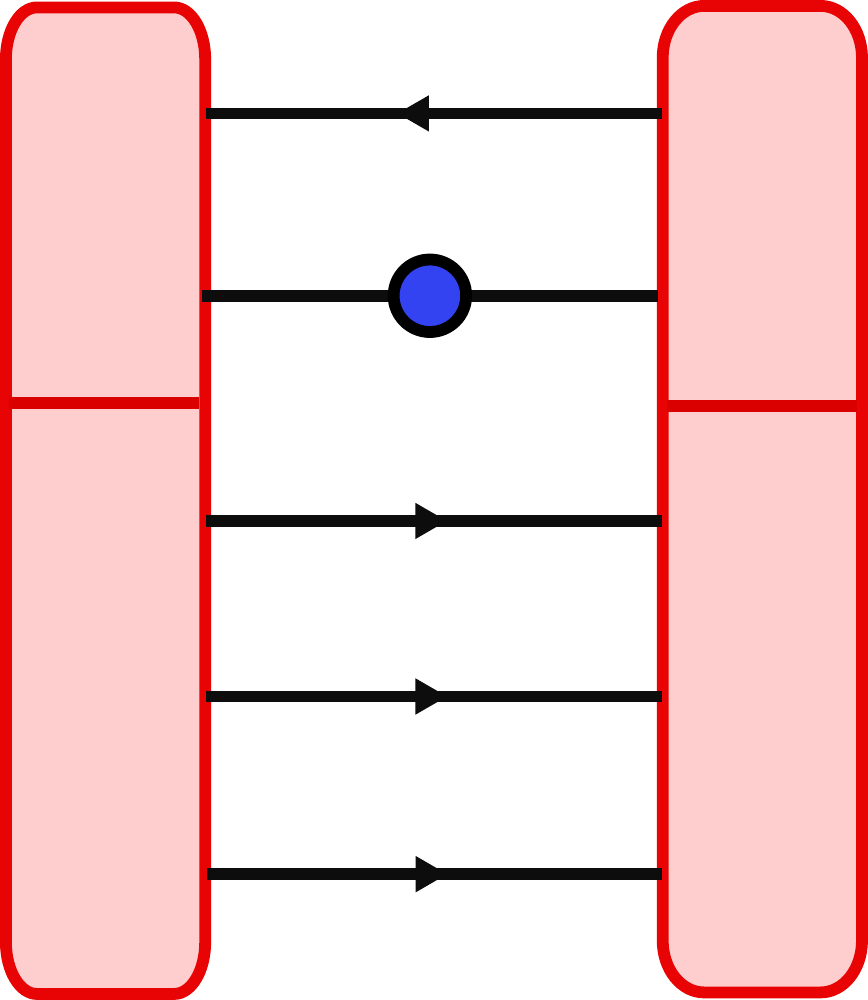} \hspace{0.1cm}
\includegraphics[scale=0.175]{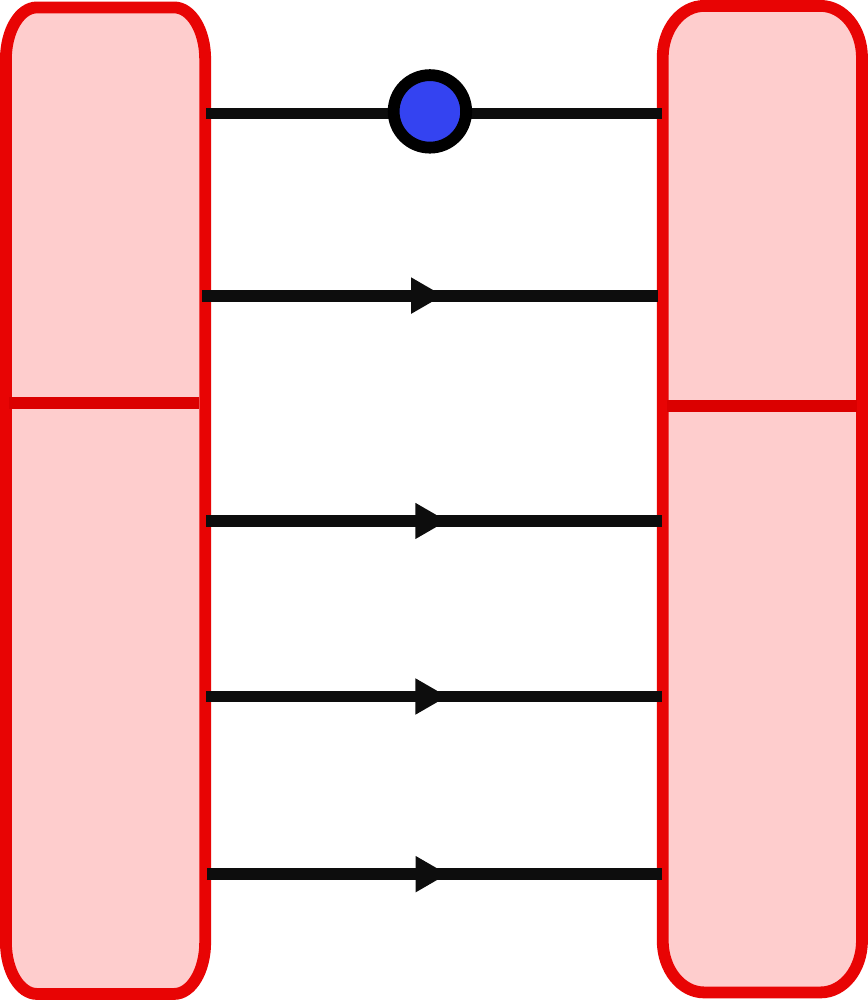}}
\caption{
The overlap representations of the $5Q$ wave functions for the quark distribution are drawn diagrammatically. The quark (right-pointing) and antiquark (left-pointing) lines are indicated by black arrows, and the non-local operator acting on these quarks is denoted by a blue circle. The valence quark contribution $f^{q_{\mathrm{val}}}$ ($\mathcal{K}_{J}$), the sea quark contribution $f^{q_{\mathrm{s}}}$ ($\mathcal{K}^{+}_{J}$), and the antiquark contribution $f^{\bar{q}}$ ($\mathcal{K}^{-}_{J}$) are shown in the left, middle, and right diagrams, respectively.}
\label{fig:1_d}
\end{figure}

{\it $5Q$ component.} Having sandwiched the partonic operator~\eqref{eq:parton} between the $5Q$ baryon wave functions~\eqref{eq:LCWF_master} and performed the contractions of the creation and annihilation operators, we derive the overlap representation for the quark distribution:
\begin{widetext}
\begin{align}
O^{q,(5)}_{BB'}(x) &=  18 T(B)_{f^{\prime}_1f_2f_3f^{\prime}_4,j^{\prime}_5}^{j^{\prime}_{1}j^{\prime}_{2}j^{\prime}_{3}j^{\prime}_{4},f^{\prime}_5,l} T(B)^{f_1f_2f_3f_4,j_5}_{j_{1}j_{2}j_{3}j_{4},f_{5},k}      \cr
&\times \int\left[d\bm{p}\right]_{5}  F^{j_1\sigma_1}(\bm{p}_1)F^{j_2\sigma_2}(\bm{p}_2)F^{j_3\sigma_3}(\bm{p}_3) W^{j_{4}\sigma_{4}}_{j_{5}\sigma_{5}}(\bm{p}_{4},\bm{p}_{5})  F^{\dagger}_{j^{\prime}_1\sigma_1}(\bm{p}_1)F^{\dagger}_{j^{\prime}_2\sigma_2}(\bm{p}_2)F^{\dagger}_{j^{\prime}_3\sigma_3}(\bm{p}_3)W^{j^{\prime}_{5}\sigma_{5}}_{cj^{\prime}_{4}\sigma_{4}}(\bm{p}_{4},\bm{p}_{5}),  \cr
&\times\bigg{[}3\delta(x_{1}-x) \delta^{f^{\prime}_{1}}_{q} \delta^{q}_{f_{1}}\delta^{f^{\prime}_{4}}_{f_{4}}\delta^{f_{5}}_{f^{\prime}_{5}} + \delta(x_{4}-x) \delta^{f^{\prime}_{4}}_{q} \delta^{q}_{f_{4}}\delta^{f^{\prime}_{1}}_{f_{1}}\delta^{f_{5}}_{f^{\prime}_{5}}  - \delta(x_{5}+x) \delta^{f^{\prime}_{5}}_{q} \delta^{q}_{f_{5}} \delta^{f^{\prime}_{1}}_{f_{1}} \delta^{f^{\prime}_{4}}_{f_{4}}\bigg{]}, 
\label{eq:UPDFs5Q}
\end{align}
\end{widetext}
To be consistent with the normalization~\eqref{eq:5Qnor}, we only consider the ``direct" quark exchange contributions in this work. While the $a^{\dagger}a$ part of the external quark operator is applied to the $3Q$ (valence quark) or the quark in the $Q\bar{Q}$ pair (sea quark), the $b^{\dagger}b$ part is applied to the antiquark in the $Q\bar{Q}$ pair. Depending on which quark is activated, the Dirac delta function $\delta(x - x_i)$ appears in the integrand~\eqref{eq:UPDFs5Q}. These contractions are shown diagrammatically in Fig.~\ref{fig:1_d}.

The number of cases for each diagram in Fig.~\ref{fig:1_d} is determined as follows:
i) The number of cases in which the external quark operator is applied to a valence quark is $3 \cdot 3 \cdot 2$, while the direct exchange of the inactive quark/antiquark between the $Q\bar{Q}$ pairs (color singlet) yields a color factor of $N_{c}$. As a result, when the valence quark becomes active, the combinatorial factor for the corresponding contractions is $54$.
ii) When the external quark operator is applied to the quark in the $Q\bar{Q}$ pair, the direct exchange with the valence quarks gives a factor of $3!$, and the direct exchange of the $Q\bar{Q}$ pair contributes a color factor of $N_{c}$. In total, the contractions give $18$.
iii) The $b^{\dagger} b$ part of the external quark operator can be applied to the antiquark in the $Q\bar{Q}$ pair. In this case, everything proceeds identically to the case where the quark in the $Q\bar{Q}$ pair is active, except that the longitudinal momentum fraction $x$ lies in the negative domain and an additional overall negative sign appears.

Before presenting the results on the quark distributions, we first discuss the dynamical parameters, which parameterize the quark distribution in a simple form.

First, when the valence quark is activated by the external quark operator, the corresponding quark distribution is expressed in terms of a linear combination of the dynamical parameters $\mathcal{K}_{J}$ ($J = \pi\pi$, $\sigma\sigma$, $33$), defined as
\begin{align}
\mathcal{K}_{J}(x) &= \frac{M^{2}}{2\pi} \int \frac{d^{3}\bm{q}}{(2\pi)^{3}} \theta(q_{z}) q_{z}  \cr
&\times \phi\left(x,\frac{q_{z}}{M_{N}},\bm{q}_{\perp}\right) G_{J}(q_{z},\bm{q}_{\perp}).
\label{eq:5Q1}
\end{align}
The generalized probability distribution $\phi(x)$ and the contributions from the pair wave function $G_{J}$ are defined in Eq.~\eqref{eq:valence_prob_2} and Eq.~\eqref{eq:dy}, respectively. By performing the integral of $\mathcal{K}_{J}(x)$ over $x$ [or using the relation~\eqref{eq:nor_pdf}], one can clearly see that Eq.~\eqref{eq:5Q1} is normalized to the $5Q$ normalization parameter $K_{J}$~\eqref{eq:nor_para}, i.e.,
\begin{align}
\int^{1}_{0} \mathcal{K}_{J}(x) = {K}_{J}.
\label{normalization:1}
\end{align}

Second, when a quark (antiquark) from the $Q\bar{Q}$ pair is activated by the external $a a^{\dagger}$ ($b b^{\dagger}$) operator, the corresponding quark distribution is expressed in terms of a linear combination of the dynamical parameters $\mathcal{K}^{+}_{J}$ ($\mathcal{K}^{-}_{J}$), defined as
\begin{align}
\mathcal{K}^{\pm}_{J}(x) &= \frac{M^{2}}{2\pi} \int \frac{d^{3}\bm{q}}{(2\pi)^{3}} \theta(q_{z}) q_{z} \cr
&\times \Phi\left(\frac{q_{z}}{M_{N}},\bm{q}_{\perp}\right)    G^{\pm}_{J}(x,q_{z},\bm{q}_{\perp}).
\label{eq:5Q2}
\end{align}
While the probability distributions are the same as those used in the normalization~\eqref{eq:dy}, the contribution of the pair wave function is generalized to $G^{\pm}_{J}(x,q_{z},\bm{q}_{\perp})$. Using the relations $x_{4}=z(1-y)$ and $x_{5}=zy$, we define $G^{\pm}_{J}(x,q_{z},\bm{q}_{\perp})$ as
\begin{align}
&\left\{\begin{array}{c} G^{+}_{J}(x,q_{z},\bm{q}_{\perp}) \\[2ex] G^{-}_{J}(x,q_{z},\bm{q}_{\perp}) \end{array} \right\} = \int^{1}_{0} dy  \int \frac{d^{2}\bm{\mathcal{Q}}_{\perp}}{(2\pi)^{2}}  \nonumber \\[1ex]
&\times \left\{\begin{array}{c} \delta(x- z(1-y)) \\[1ex] -\delta(x+ zy) \end{array} \right\}  \mathcal{G}_{J} (\bm{q},y,\bm{\mathcal{Q}}_{\perp})   \bigg{|}_{\mathrm{reg}}.
\label{eq:5Q2_2}
\end{align}
Since $G^{\pm}_J$ are logarithmically divergent quantities, we introduce the Pauli–Villars regulator to tame them [cf.~\eqref{eq:reg}].
Here, the integral of the Dirac delta function over $y$ gives a step function. Thus, one can clearly see that the $x$ domain for the quark and antiquark distributions lies in the positive and negative regions, respectively.

Importantly, one can observe the following symmetry between the dynamical parameters for the active quark and active antiquark in the $Q\bar{Q}$ pair:
\begin{align}
\mathcal{K}^{+}_{J}(x)= -\mathcal{K}^{-}_{J}(-x).
\label{sy}
\end{align}
This can be easily verified by changing the variable $1 - y \to y'$ in Eq.~\eqref{eq:5Q2_2}. Here, we used the fact that the functions $\mathcal{G}_{J}$ in Eq.~\eqref{eq:density} are invariant under this change of variables.

Having integrated Eq.~\eqref{eq:5Q2_2} over $x$, we observe that the pair wave function contribution in Eq.~\eqref{eq:5Q2_2} becomes the one used in normalization [cf.\eqref{eq:dy}]:
\begin{align}
 \int^{1}_{-1} dx \,  G^{\pm}_{J}(x,q_{z},\bm{q}_{\perp})= \pm G_{J}(q_{z},\bm{q}_{\perp}).
\end{align}
This result implies that the integral of Eq.~\eqref{eq:5Q2} over $x$ recovers the dynamical parameters used in the normalization~\eqref{eq:nor_para}:
\begin{align}
& \int^{1}_{-1} dx \, \mathcal{K}^{\pm}_{J}(x)= \pm K_{J}.
\label{normalization:2}
\end{align}

After summing over all the spin/isospin/flavor indices in the overlap representation~\eqref{eq:UPDFs5Q}, we now derive the quark distributions for the proton in terms of the dynamical functions in Eqs.~\eqref{eq:5Q1} and \eqref{eq:5Q2} (see Appendix~\ref{appendix:b} for the results for the $\Delta^{+}$ baryon):
\begin{itemize}
\item Active $3Q$ and inactive $Q\bar{Q}$ pair (left diagram in Fig.~\ref{fig:1_d})
\begin{align}
O^{u_{\mathrm{val}},(5)}_{p}(x)&=\frac{3}{50} \left[ 225\mathcal{K}_{\sigma \sigma}(x) + 89\mathcal{K}_{\pi \pi} (x) \right], \cr
O^{d_{\mathrm{val}},(5)}_{p}(x)&=\frac{3}{25} \left[ 55\mathcal{K}_{\sigma \sigma}(x) + 31\mathcal{K}_{\pi \pi} (x) \right], \cr
O^{s_{\mathrm{val}},(5)}_{p}(x)&=\frac{3}{25} \left[ 5\mathcal{K}_{\sigma \sigma}(x) + 7\mathcal{K}_{\pi \pi} (x) \right].
\end{align} 
\item Inactive $3Q$ and active quark in the $Q\bar{Q}$ pair (middle diagram in Fig.~\ref{fig:1_d})
\begin{align}
O^{u_{\mathrm{s}},(5)}_{p}(x)&=\frac{3}{50} \left[ 53\mathcal{K}^{+}_{\sigma \sigma}(x) + 37\mathcal{K}^{+}_{\pi \pi} (x) \right], \cr
O^{d_{\mathrm{s}},(5)}_{p}(x)&=\frac{1}{50} \left[ 137\mathcal{K}^{+}_{\sigma \sigma}(x) + 53\mathcal{K}^{+}_{\pi \pi} (x) \right], \cr
O^{s_{\mathrm{s}},(5)}_{p}(x)&=\frac{1}{50} \left[ 49\mathcal{K}^{+}_{\sigma \sigma}(x) + \mathcal{K}^{+}_{\pi \pi} (x) \right].
\end{align} 
\item Inactive $3Q$ and active antiquark in the $Q\bar{Q}$ pair (right diagram in Fig.~\ref{fig:1_d})
\begin{align}
O^{\bar{u},(5)}_{p}(x)&=\frac{24}{25} \left[ 3\mathcal{K}^{-}_{\sigma \sigma}(x) + \mathcal{K}^{-}_{\pi \pi} (x) \right], \cr
O^{\bar{d},(5)}_{p}(x)&=\frac{1}{25} \left[ 61\mathcal{K}^{-}_{\sigma \sigma}(x) + 37\mathcal{K}^{-}_{\pi \pi} (x) \right], \cr
O^{\bar{s},(5)}_{p}(x)&=\frac{1}{50} \left[ 79\mathcal{K}^{-}_{\sigma \sigma}(x) + 43\mathcal{K}^{-}_{\pi \pi} (x) \right].
\end{align} 
\end{itemize}
Comparing the results for the $5Q$ wave function with those for the $3Q$ wave function, we find not only a nonzero strange quark contribution, but also nonzero sea and antiquark contributions. In particular, the contributions of the $5Q$ wave function to the proton quark distributions are characterized solely by the four independent dynamical parameters:
\begin{align}
\mathcal{K}_{\sigma \sigma, \pi \pi}(x), \quad \mathcal{K}^{+}_{\sigma \sigma, \pi \pi}(x)= -\mathcal{K}^{-}_{\sigma \sigma, \pi \pi}(-x).
\end{align}

Now, we have computed both the contributions of the $3Q$ and $5Q$ components of the wave functions to the quark distributions. By combining these contributions and normalizing the wave functions, we obtain the physical quark distributions, defined as
\begin{align}
&f^{q}_{B B'}(x) \cr
&= \frac{O^{q, (3)}_{B B'}(x)+O^{q, (5)}_{B B'}(x)}{\sqrt{\mathcal{N}^{(3)}(B)+\mathcal{N}^{(5)}(B)} \sqrt{\mathcal{N}^{(3)}(B')+\mathcal{N}^{(5)}(B')}},
\label{eq:nor22}
\end{align}
where the contributions of the $5Q$ wave function to the quark distribution appear as corrections to those of the $3Q$ wave function. Using the fact that $\mathcal{K}_{J}(x)$ and $\mathcal{K}^{\pm}_{J}(x)$ are normalized to the dynamical parameters used in the wave function normalizations [cf.~\eqref{normalization:1} and \eqref{normalization:2}], we clearly see that the following sum rules are satisfied for the proton:
\begin{align}
&\int^{1}_{-1} dx \, f^{u}_{p}(x) = 2, \quad \int^{1}_{-1} dx \, f^{d}_{p}(x) = 1, \cr
&\int^{1}_{-1} dx \, f^{s}_{p}(x) = 0.
\end{align}
Moreover, it is straightforward to verify that the integral of the $x$-weighted quark distributions over $x$ yields the correct momentum sum rule:
\begin{align}
\int^{1}_{-1} dx \, x f^{u+d+s}_{p}(x) = 1.
\end{align}

Similarly, we also derive the quark distributions for the $p\to \Delta^{+}$ transition:
\begin{itemize}
    \item Active $3Q$ and inactive $Q\bar{Q}$ pair (left diagram in Fig.~\ref{fig:1_d})
\begin{align}
O^{u_{\mathrm{val}},(5)}_{p\Delta^{+}}(x) &=-O^{d_{\mathrm{val}},(5)}_{p\Delta^{+}}(x) \cr
&= \frac{3}{5 \sqrt{5}}  \left[\mathcal{K}_{\pi \pi}(x)-3 \mathcal{K}_{33}(x) \right], \cr
O^{s_{\mathrm{val}},(5)}_{p\Delta^{+}}(x) &= 0. 
\label{eq:t_1}
\end{align}
\item Inactive $3Q$ and active quark in the $Q\bar{Q}$ pair (middle diagram in Fig.~\ref{fig:1_d})
\begin{align}
O^{u_{\mathrm{s}},(5)}_{p\Delta^{+}}(x) &=-O^{d_{\mathrm{s}},(5)}_{p\Delta^{+}}(x) \cr
&=  -\frac{3}{10 \sqrt{5}}  \left[\mathcal{K}^{+}_{\pi \pi}(x)-3 \mathcal{K}^{+}_{33}(x)\right], \cr
O^{s_{\mathrm{s}},(5)}_{p\Delta^{+}}(x) &= 0.
\label{eq:t_2}
\end{align}
\item Inactive $3Q$ and active antiquark in the $Q\bar{Q}$ pair (right diagram in Fig.~\ref{fig:1_d})
\begin{align}
O^{\bar{u},(5)}_{p\Delta^{+}}(x) &=-O^{\bar{d},(5)}_{p\Delta^{+}}(x) \cr
&= \frac{3}{10 \sqrt{5}}  \left[\mathcal{K}^{-}_{\pi \pi}(x)-3 \mathcal{K}^{-}_{33}(x) \right], \cr
O^{\bar{s},(5)}_{p\Delta^{+}}(x) &= 0.
\label{eq:t_3}
\end{align}
\end{itemize}
Using Eq.~\eqref{eq:nor22}, we construct the physical quark distribution function $f^{q}_{p\Delta^{+}}(x)$ in the $p\to \Delta^{+}$ transition. We have already shown that there is no contribution from the $3Q$ wave function to the $p \to \Delta^{+}$ quark distribution. This means that the leading contribution appears only in the overlap representation of the $5Q$ wave function.

We find that the $u$-quark and $d$-quark contributions have opposite signs, and that there are no $s$ or $\bar{s}$ contributions. Thus, we conclude that the quark distribution for the flavor singlet/octet component vanishes, while the flavor triplet component remains finite.

Importantly, the quark distribution is characterized by the dynamical parameters $\mathcal{K}^{(\pm)}_{\pi \pi} - 3 \mathcal{K}^{(\pm)}_{33}$, which are interpreted as the quadrupole deformation of the pion cloud:
\begin{align}
&\mathcal{K}^{(\pm)}_{\pi \pi}(x) - 3\mathcal{K}^{(\pm)}_{33}(x) \nonumber \\[1ex]
&\propto \int \frac{d^{3}\bm{q}}{(2\pi)^{3}} \left(1 - 3\frac{q^{2}_{z}}{|\bm{q}|^{2}}\right) \Pi^2 ({\bm{q}}) [...]. 
\end{align}
This structure exhibits the standard quadrupole deformation $\propto Q^{33}$ in momentum space. Notably, the $\Sigma$ exchange does not contribute, since a scalar exchange cannot generate a quadrupole structure from the QCD vacuum.

Finally, we take the first $x$-moment of the quark distribution and find
\begin{align}
\int^{1}_{-1} dx \, f^{q}_{p\Delta^{+}}(x) = 0.
\label{eq:sr_tr}
\end{align}
In Sec.~\ref{sec:7}, we will verify this result more generally in the context of the transition GPDs.

\section{Numerical results \label{sec:6}}
Before discussing the results for the quark distribution, we first explain how the numerical values of the dynamical parameters are determined. The mean-field solution $U_{\mathrm{cl}}(\bm{x})$ was obtained in Ref.~\cite{Diakonov:1988mg} in a self-consistent way, and the corresponding result is well approximated by the given arctangent profile function
\begin{align}
P(r) = 2 \arctan\left(\frac{R^{2}_{0}}{r^{2}}\right), \quad M R_{0}= 0.8,
\label{eq:profile2}
\end{align}
where $R_{0}$ is the average size of the mean field. As already explained in Sec.~\ref{sec:4}, the dynamical quark mass $M$ in Eq.~\eqref{eq:profile2} is originally momentum-dependent, originating from the quark zero-mode form factor induced by the instanton. Using the standard parameters for the instanton ensemble, one obtains $M = M(0) = 345$~MeV at zero quark virtuality. However, for simplicity, we turn off the momentum dependence. This means that we have to introduce the Pauli-Villars~(PV) cutoff mass to tame the UV divergence. The PV cutoff mass is set to $M_{\mathrm{PV}} = 557$~MeV by reproducing the pion decay constant $f_{\pi} = 93$~MeV. Using the self-consistent profile function~\eqref{eq:profile2} and employing the PV cutoff mass, one obtains the classical soliton mass as $M_{N} = 1.207~\mathrm{GeV}$; see Ref.~\cite{Diakonov:1988mg}.

With the fixed parameters above, we compute the numerical values of the dynamical parameters $K_{J}$~($J=\sigma \sigma, \pi \pi, 33$) used in the wave function normalization. They are estimated as follows:
\begin{align}
K_{\sigma \sigma}=0.029, \quad K_{\pi \pi}=0.075, \quad K_{3 3}=0.041.
\label{eq:dy_nu}
\end{align}
Using these numerical values, we determine the normalizations for both the baryon octet and decuplet:
\begin{align}
&\mathcal{N}^{(5)}(B_{8}) = 0.45, \cr
&\mathcal{N}^{(5)}_{1/2}(B_{10}) = 0.23, \quad \mathcal{N}^{(5)}_{3/2}(B_{10}) = 0.21.
\end{align}

\begin{figure}[htp]
\centerline{%
\includegraphics[scale=0.28]{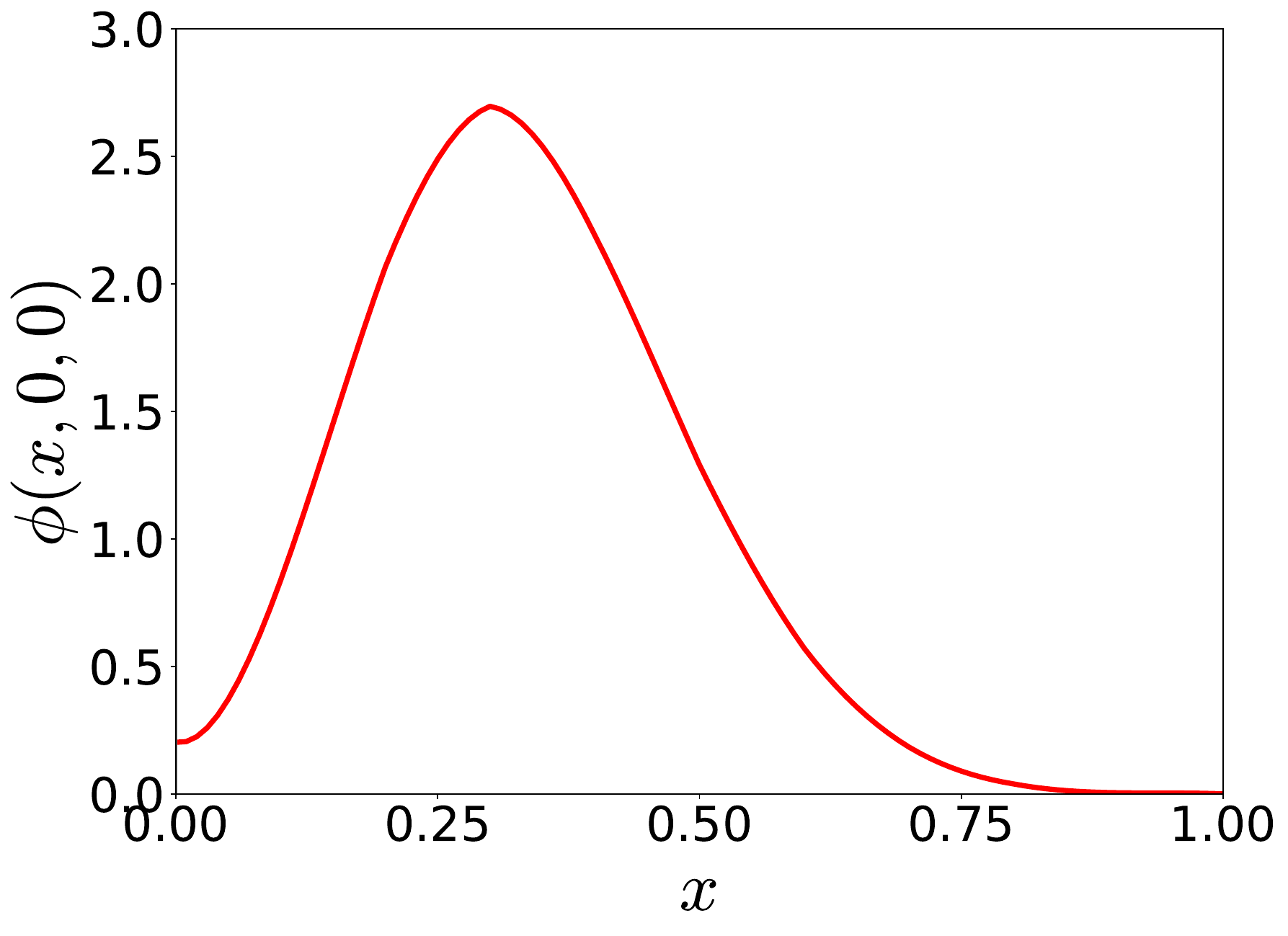}}
\caption{The generalized probability distribution $\phi(x,0,0)$ is plotted as a function of $x$.}
\label{fig:2}
\end{figure}

In Sec.~\ref{sec:5_2}, we show that the contribution of the $3Q$ wave function to the quark distribution, $f^{q_\mathrm{val},(3)}$, is expressed in terms of the generalized probability distribution $\phi(x,0,0)$. In Fig.~\ref{fig:2}, we plot $\phi(x,0,0)$ as a function of $x$. The peak of $\phi(x,0,0)$ is located around $x \approx 0.3$. Since $\Phi(0,0) = 1$, the integral of $\phi(x,0,0)$ over $x$ is numerically normalized to unity.

\begin{figure}[htp]
\centerline{%
\includegraphics[scale=0.28]{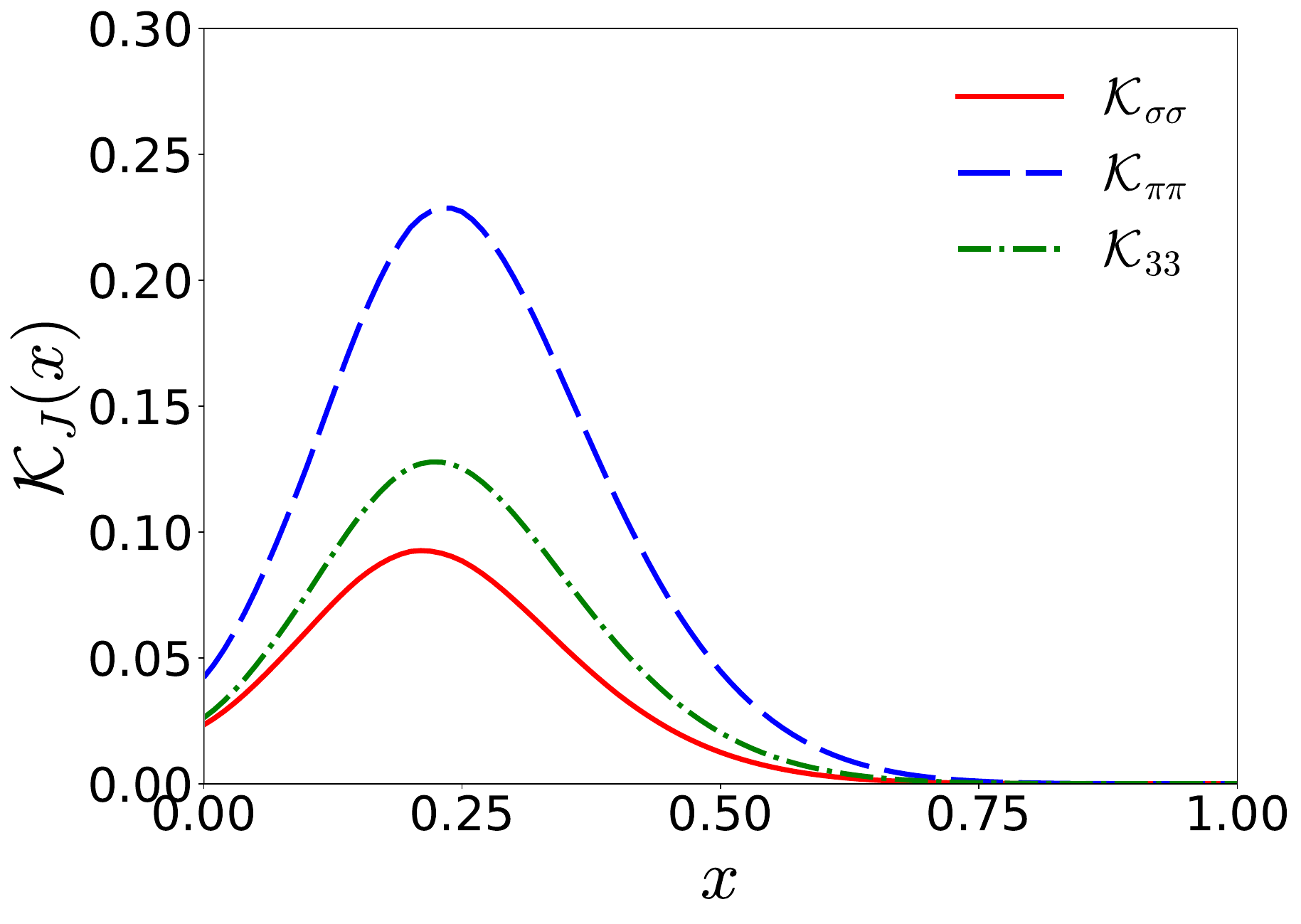}}
\caption{The dynamical parameters $\mathcal{K}_{J}(x)$, arising from the overlap of the $5Q$ wave function, are plotted as functions of $x$.}
\label{fig:3}
\end{figure}
In Fig.~\ref{fig:3}, we show the dynamical parameters $\mathcal{K}_{J}(x)$ that characterize the valence quark distribution $f^{q_\mathrm{val},(5)}$ resulting from the overlap of the $5Q$ wave functions. Similar to Fig.~\ref{fig:2}, they exhibit a typical valence distribution shape with a peak around $x \approx \frac{1}{3}$. However, compared to the valence quark contribution from the $3Q$ wave function (Fig.~\ref{fig:2}), these distributions are more concentrated at smaller $x$ values.

Depending on the exchanged meson (or isospin quantum number), the dynamical parameters are classified into the pion contributions ($\mathcal{K}_{\pi \pi}, \mathcal{K}_{33}$) and the sigma meson contribution ($\mathcal{K}_{\sigma \sigma}$). We find that, in general, the pion exchange channel dominates over the scalar exchange channel in Fig.~\ref{fig:3}, a feature that can be quantitatively understood from Eq.~\eqref{eq:dy_nu} via the normalization condition in Eq.~\eqref{normalization:1}.

\begin{figure}[htp]
\centerline{%
\includegraphics[scale=0.28]{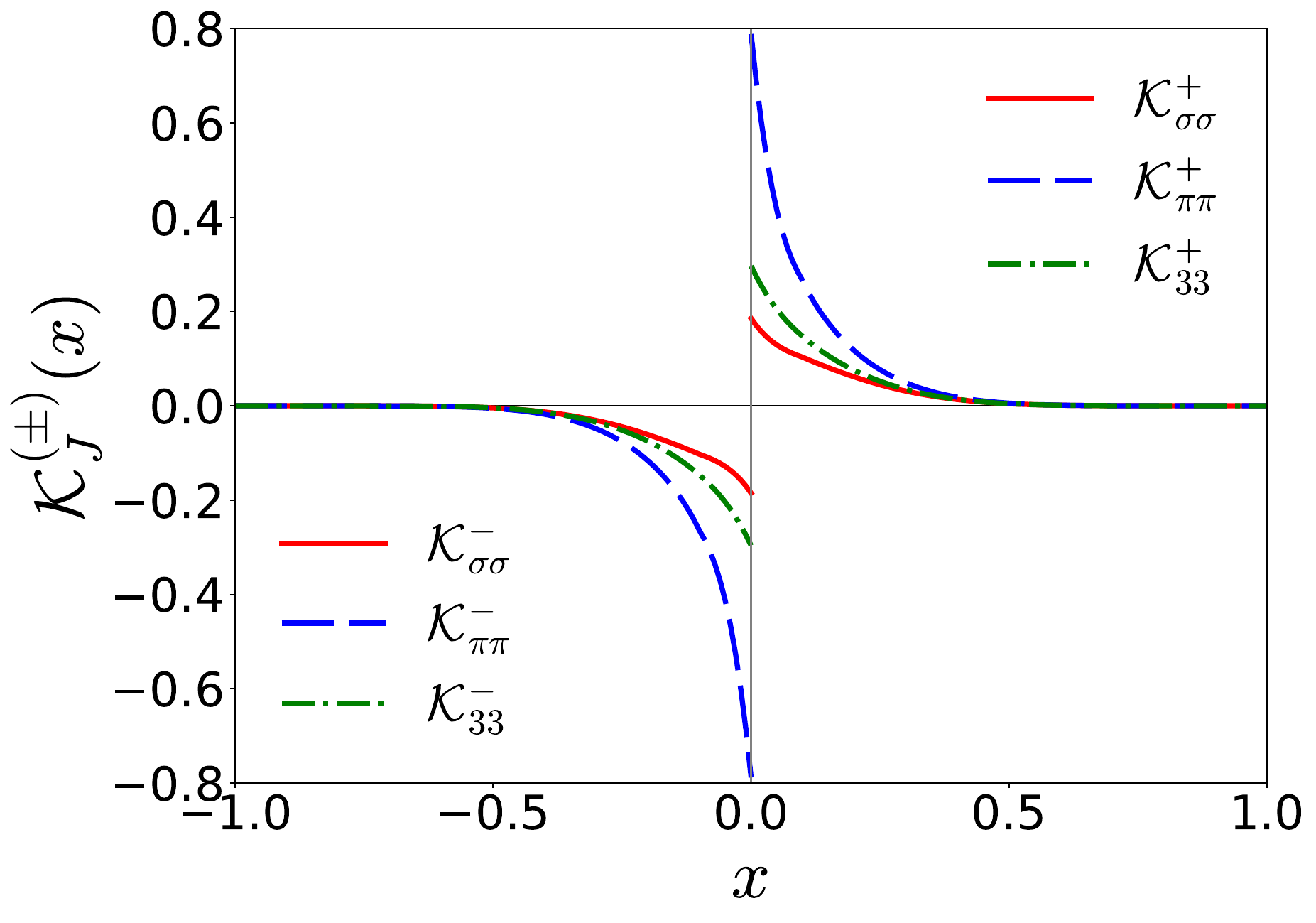}}
\caption{The dynamical parameters $\mathcal{K}^{\pm}_{J}(x)$, arising from the overlap of the $5Q$ wave function, are plotted as functions of $x$.}
\label{fig:4}
\end{figure}
In Fig.~\ref{fig:4}, we plot the dynamical parameters $\mathcal{K}^{\pm}_{J}$ that characterize the sea quark $f^{q_\mathrm{s},(5)}$ and antiquark $f^{\bar{q},(5)}$ contributions to the quark distributions arising from the overlap of the $5Q$ wave function. The domains of the dynamical parameters $\mathcal{K}^{+}_{J}$ (active quark) and $\mathcal{K}^{-}_{J}$ (active antiquark) extend over $x=[0,1]$ and $x=[-1,0]$, respectively. Furthermore, we confirm that the quark-antiquark pair symmetry~\eqref{sy} is numerically satisfied. These dynamical parameters mainly govern the small-$x$ behavior (or chiral enhancement) of the quark distribution. In particular, we observe that the chiral enhancement is largely due to the contribution from pion exchange ($\mathcal{K}^{\pm}_{\pi \pi}, \mathcal{K}^{\pm}_{33}$).

\begin{figure}[htp]
\centerline{%
\includegraphics[scale=0.28]{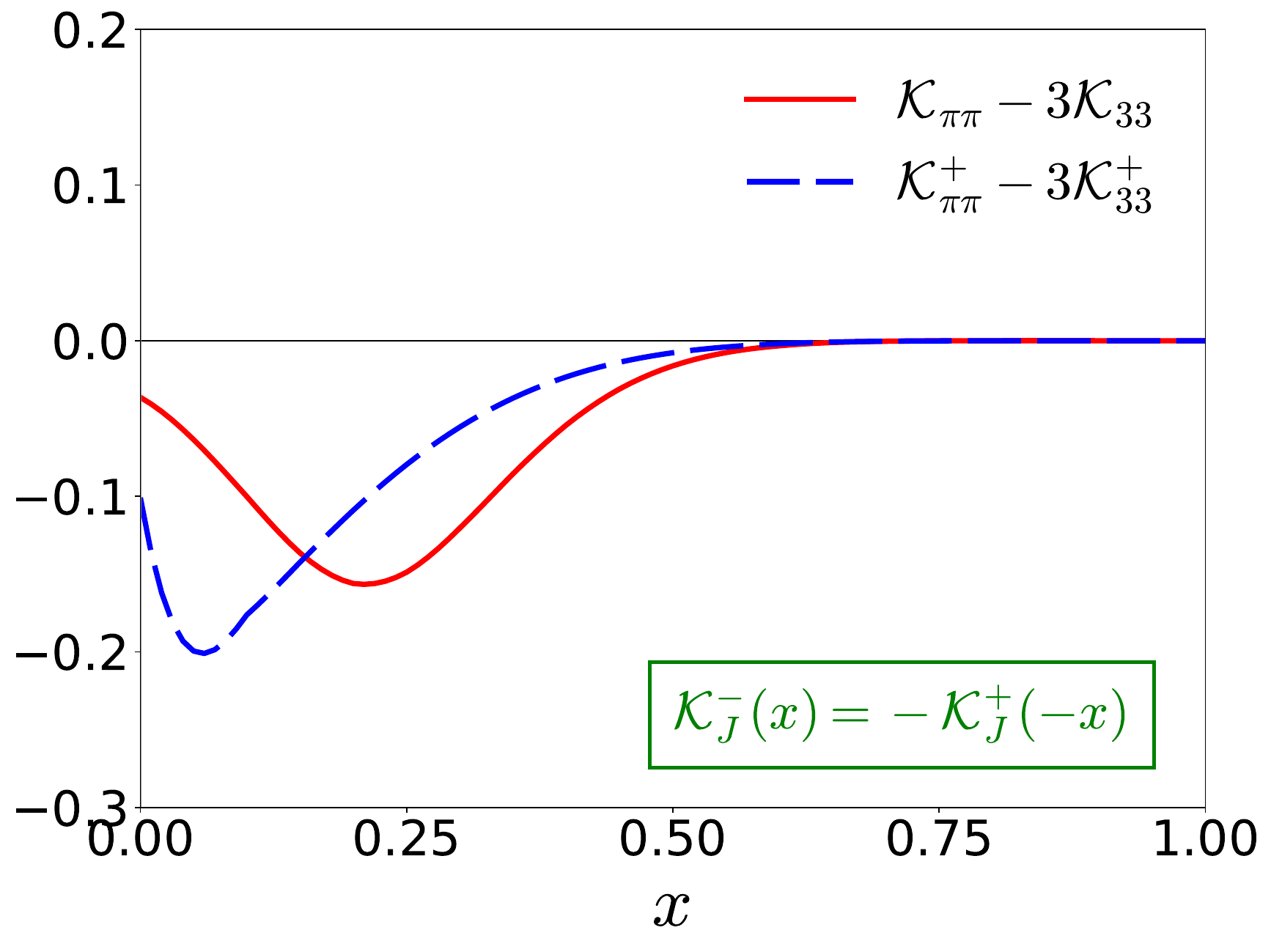}}
\caption{The ``quadrupole" dynamical parameters $\mathcal{K}_{\pi \pi}(x) - 3\mathcal{K}_{33}(x)$, resulting from the overlap of the $5Q$ proton and $\Delta^{+}$ baryon wave functions, are plotted as a function of $x$.}
\label{fig:5}
\end{figure}
Using the dynamical parameters obtained above, we are now able to determine the quark distribution for the proton, the $\Delta^{+}$ baryon, and the $p \to \Delta^{+}$ transition. In this work, we will focus on the numerical results for the quark distributions in the $p \to \Delta^{+}$ transition. As mentioned in the previous section, we see that the leading contribution to the $p \to \Delta^{+}$ quark distribution comes from the overlap of the $5Q$ wave function. This contribution is characterized by $\mathcal{K}^{(\pm)}_{\pi \pi} - 3 \mathcal{K}^{(\pm)}_{33}$ and is interpreted as a quadrupole deformation of the pion cloud. These ``quadrupole'' dynamical parameters are drawn in Fig.~\ref{fig:5}. As shown in Fig.~\ref{fig:4}, the two dynamical parameters $\mathcal{K}^{+}_{\pi \pi}$ and $3 \mathcal{K}^{+}_{33}$ almost cancel each other. This strong cancellation results in a small sea quark contribution with a bump at around $x \sim 0.1$. Since the magnitude of $-3 \mathcal{K}^{+}_{33}$ is larger than that of $\mathcal{K}^{+}_{\pi \pi}$, we obtain a negative sea quark contribution, i.e., $\mathcal{K}^{+}_{\pi \pi} - 3 \mathcal{K}^{+}_{33} < 0$. Similarly, the valence quark contribution becomes very small and negative. Using the quark–antiquark pair symmetry~\eqref{sy}, the antiquark contribution can also be simply determined.

\begin{figure}[htp]
\centerline{%
\includegraphics[scale=0.28]{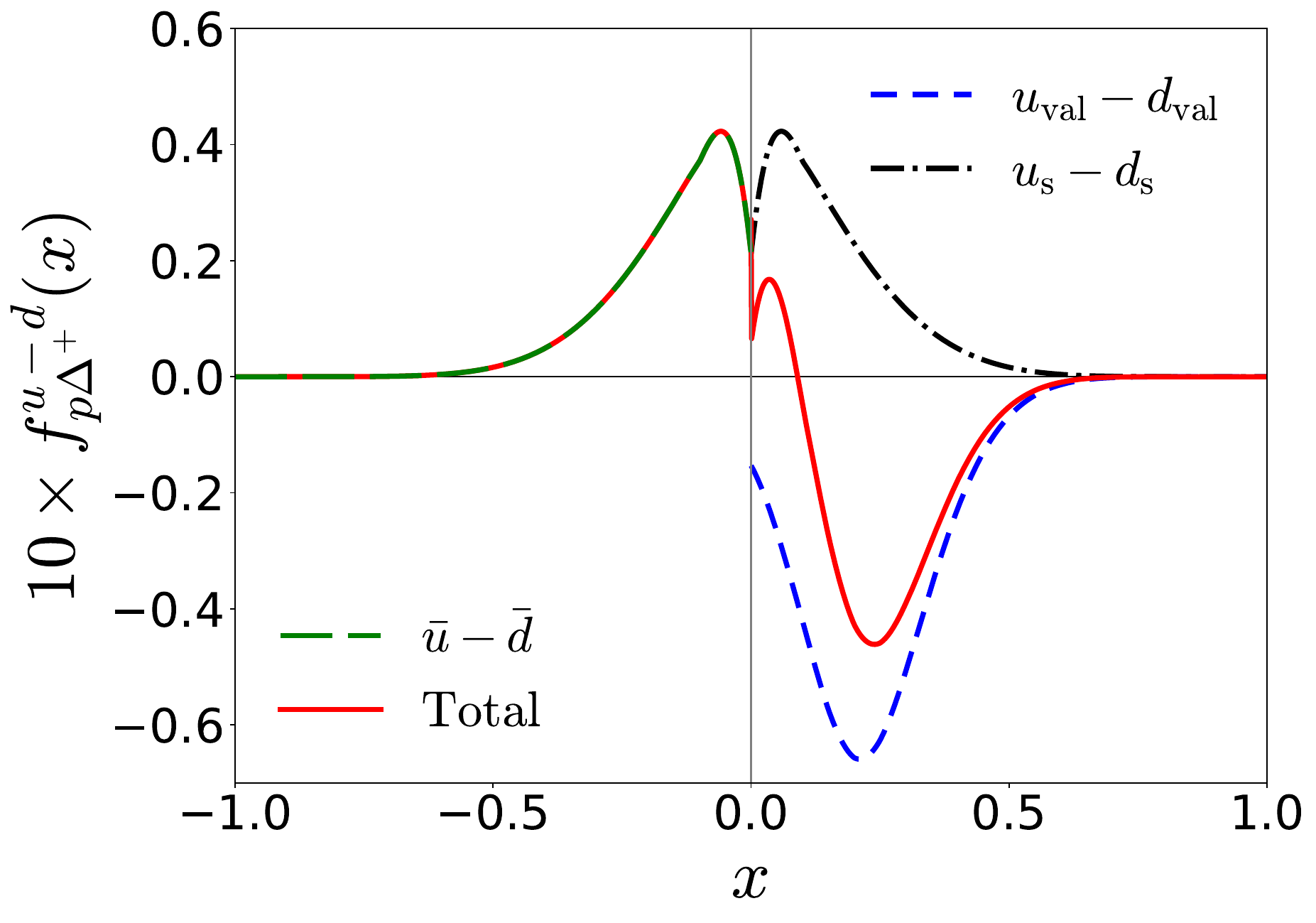}}
\caption{The quark distribution in the $p \to \Delta^{+}$ transition is drawn as a function of $x$.
{\it Solid (red) curve}: total.
{\it Long-dashed (green) curve}: antiquark contribution.
{\it Short-dashed (blue) curve}: valence quark contribution.
{\it Dot-dashed (black) curve}: sea quark contribution.}
\label{fig:6}
\end{figure}
By combining the valence and sea quark contributions, we draw the isovector component of the quark distributions in the $p \to \Delta^{+}$ transition in Fig.~\ref{fig:6}. The antiquark and sea quark contributions reflect the quark–antiquark pair symmetry~\eqref{sy}, are positive over $x = [-1,1]$, and have peaks at around $|x| \sim 0.1$. On the other hand, the valence quark contribution is negative over $x = [0,1]$ and has a peak near $x \approx 1/3$. In total, there are three peaks and one nodal point over $x = [-1,1]$. As derived in Eq.~\eqref{eq:sr_tr}, we numerically confirm that the first moment of the quark distribution vanishes:
\begin{align}
\int^{1}_{-1} dx f^{u-d}_{p\Delta^{+}}(x)& = 0.
\end{align}
When it comes to the second moment of the quark distribution, there is no conserved quantity, so we get the following non-trivial result:
\begin{align}
\int^{1}_{-1} dx \, x f^{u-d}_{p\Delta^{+}}(x)& \approx -0.004.
\end{align}
We obtain a remarkably small number, which is consistent with the standard large-$N_{c}$ prediction~\cite{Goeke:2001tz, Kim:2024hhd, Kim:2025ilc}. The values of the first and second $x$-moments of the quark distribution for the separate valence and sea quark contributions are listed in Tab.~\ref{tab:1}.
\begin{table}[htp]
\centering
\setlength{\tabcolsep}{3pt}
\renewcommand{\arraystretch}{1.5}
\caption{The values of the first and second $x$-moments of the $p \to \Delta^{+}$ unpolarized quark distribution for the separate valence, sea quark, and antiquark contributions are listed. }
\begin{tabular}{ c |  c  c  c   | c c c  } 
\hline
\hline
 Contents &  \multicolumn{3}{c |}{$u$} & \multicolumn{3}{c }{$d$}  \\
 \hline
 $\int dx \, f^{q}_{p\Delta^{+}} $   & $q_{\mathrm{val}}$ & $q_{\mathrm{s}}$ & $\bar{q}$ & $q_{\mathrm{val}}$ & $q_{\mathrm{s}}$ & $\bar{q}$  \\
\hline
 $3Q$    & $0$ & $0$ &  $0$& $0$& $0$& $0$ \\
 $3Q+5Q$ & $-0.010$ & $0.005$ & $0.005$ & $0.010$ & $-0.005$ & $-0.005$  \\
\hline
 Contents &  \multicolumn{3}{c |}{$u$} & \multicolumn{3}{c }{$d$}  \\
 \hline
 $\int dx \, x f^{q}_{p\Delta^{+}} $   & $q_{\mathrm{val}}$ & $q_{\mathrm{s}}$ & $\bar{q}$ & $q_{\mathrm{val}}$ & $q_{\mathrm{s}}$ & $\bar{q}$  \\
\hline
 $3Q$    & $0$ & $0$ &  $0$& $0$& $0$& $0$ \\
 $3Q+5Q$ & $-0.002$ & $0.001$ & $-0.001$ & $0.002$ & $-0.001$ & $0.001$  \\
\hline
\hline
\end{tabular} 
\label{tab:1}
\end{table}

\section{Discussion \label{sec:7}}

\subsection{Relation between GPD and form factor}
So far, we have discussed the quark distributions defined by the forward matrix element of the QCD operator~\eqref{eq:parton}. By introducing the non-forward matrix element~\eqref{eq:1}, the quark distributions can be viewed as a part of the GPDs, which integrate the concepts of the generalized form factor and the quark distribution. In this section, we will provide the explicit relation between them. Very recently, this non-forward matrix element~\eqref{eq:1} between the proton and $\Delta^{+}$ baryon states has been parameterized in terms of a complete set of four independent GPDs~\cite{Kim:2024hhd}:
\begin{align}
\mathcal{M}_{p\Delta^{+}}[\Gamma_{q}]
&= \sqrt{\frac{2}{3}} \bar{u}_{\alpha}
\left[M_{N} n^{\alpha} \slashed{n}  \gamma_{5} H^{q}_{X}(x,\xi,t) \right] u  + ...
\label{eq:new_para}
\end{align}
where $\bar{u}_{\alpha} \equiv \bar{u}_{\alpha}(p',s')$ and $u \equiv u(p,s)$ are Rarita-Schwinger and Dirac spinors, respectively, and $H_{X}$ denotes one of the transition GPDs relevant to the unpolarized quark distribution. The coefficient $\sqrt{2/3}$ comes from the isospin factor for the $p \to \Delta^{+}$ transition. The ellipsis is irrelevant to the present discussion; see Ref.~\cite{Kim:2024hhd} for the explicit expression.

The first $x$-moment of the non-local vector current is related to the local vector current. Thus, the first $x$-moment of the matrix element~\eqref{eq:new_para} is given by
\begin{align}
\int^{1}_{-1} dx \, \mathcal{M}_{p\Delta^{+}} [\Gamma_{q}] =  n_{\mu} \mathcal{J}^{\mu}_{q},
\end{align}
where $\mathcal{J}^{\mu}_{q}$ denotes the transition matrix element of the vector current
\begin{align}
\mathcal{J}^{\mu}_{q} = \langle \Delta^{+} | \bar{\psi}_{q}(0) \gamma^{\mu} \psi_{q}(0) |  p  \rangle.
\label{eq:vector}
\end{align}
Comparing Eq.~\eqref{eq:new_para} with the parameterization of Eq.~\eqref{eq:vector} in Ref.~\cite{Kim:2024hhd}, one arrives at the vanishing first moment of the transition GPD $H_{X}$:
\begin{align}
\int^{1}_{-1} dx \, H^{q}_{X}(x,\xi,t) =  0.
\end{align}
In a general transition matrix element, this feature can be explained by invoking current conservation (see Ref.~\cite{Kim:2025ilc}).

Similarly, the second $x$-moment of the non-local vector current is related to the quark part of the symmetric EMT current. The second $x$-moment of the matrix element~\eqref{eq:new_para} is given by
\begin{align}
\int^{1}_{-1} dx \, x \mathcal{M}_{p\Delta^{+}} [\Gamma_{q}] =  n_{\mu} n_{\nu} \mathcal{T}^{\mu \nu}_{q}, 
\end{align}
where $\mathcal{T}^{\mu \nu}_{q}$ is the transition matrix element of the EMT current:
\begin{align}
\mathcal{T}^{\mu \nu}_{q} = \langle \Delta^{+}  | \bar{\psi}_{q}(0) \frac{i}{2} \overleftrightarrow{\nabla}^{ \{\mu} \gamma^{\nu\} } \psi_{q}(0) |  p \rangle.
\end{align}
Here, $\nabla$ stands for the covariant derivative, with $\overleftrightarrow{\nabla} \equiv (\overrightarrow{\nabla} - \overleftarrow{\nabla})/2$, and $a^{ \{ \mu } b^{\nu \} } = a^\mu b^{\nu} + a^\nu b^{\mu}$. In Ref.~\cite{Kim:2022bwn}, the transition matrix element of the EMT current was parameterized as follows:
\begin{align}
\mathcal{T}^{\mu \nu}_{q} = \sqrt{\frac{2}{3}} \bar{u}_{\alpha} 
\left[M_{N} \gamma^{ \{\mu} g^{\nu \} \alpha}  \gamma_{5} F^{q}_{4}(t) \right] u  + ... .
\label{eq:EMT}
\end{align}
The ellipsis denotes terms irrelevant to the present discussion.
Comparing Eq.~\eqref{eq:new_para} with Eq.~\eqref{eq:EMT}, one observes that the second $x$-moment of the GPD $H_{X}$ is related to the non-vanishing transition EMT form factor $F_{4}$ as follows:
\begin{align}
\int^{1}_{-1} dx \, x H^{q}_{X}(x,\xi,t) = 2 F^{q}_{4}(t).
\label{eq:smoment}
\end{align}

\subsection{Relations to quark distribution \label{sec:7_2}}
In order to relate the quark distribution to the GPDs and transition EMT form factors, we need to specify a frame. We adopt the 2D collinear/symmetric frame, where $\bm{P}_{\perp} = 0$ and $\bm{\Delta}_{\perp} \neq 0$. This frame is typically used in the study of GPDs. For the non-diagonal matrix element, the initial and final 4-momenta satisfy the following on-mass-shell conditions:
\begin{align}
\Delta \cdot P &= \frac{M^{2}_{\Delta}-M^{2}_{N}}{2}, \cr
 P^{2}+\frac{\Delta^{2}}{4}&=\frac{M^{2}_{N}+M^{2}_{\Delta}}{2}.
\label{eq:on_shell}
\end{align}
By employing the on-shell conditions~\eqref{eq:on_shell} and the collinear condition $\bm{P}_{\perp} = 0$, one can determine the minus components of $P$ and $\Delta$:
\begin{align}
P^{-}= \frac{ 2\xi(M^{2}_{\Delta}-M^{2}_{N})+ (2M^{2}_{\Delta}+2M^{2}_{N} +\bm{\Delta}^{2}_{\perp})}{8P^{+} (1-\xi^{2})}, \cr
\Delta^{-}= \frac{ 2(M^{2}_{\Delta}-M^{2}_{N}) + \xi (2M^{2}_{\Delta}+2M^{2}_{N} + \bm{\Delta}^{2}_{\perp})}{4P^{+}(1-\xi^{2}) },
\label{eq:GPD_frame1}
\end{align}
with $\Delta^{+} = -2\xi P^{+}$. In the forward limit ($\xi, t \to 0$), the 4-momenta $P$ and $\Delta$ are reduced to (in the notation $v = [v^{+}, v^{-}, \bm{v}_{\perp}]$):
\begin{align}
&P= \left(P^{+}, \frac{ M^{2}_{\Delta}+M^{2}_{N}}{4P^{+}}, \bm{0}_{\perp} \right), \cr
&\Delta= \left( 0, \frac{ M^{2}_{\Delta}-M^{2}_{N} }{2P^{+} }, \bm{0}_{\perp} \right).
\label{eq:GPD_frame2}
\end{align}
Note that even if we take the forward limit, we still have a non-vanishing momentum transfer, i.e., $\Delta \neq 0$. Only in the equal-mass limit can one restore the vanishing momentum transfer $\Delta = 0$; see Ref.~\cite{Kim:2025ilc} for details. 

Considering the forward limit and taking the light-front helicity states, we compute the matrix element of the spinors in Eq.~\eqref{eq:new_para} and find that it is expressed as the quadrupole spin operator:
\begin{align}
\bar{u}_{\alpha}
\sqrt{\frac{2}{3}}\left[M_{N} n^{\alpha} \slashed{n}  \gamma_{5} \right] u  = \frac{4}{3} \frac{M_{N}}{M_{\Delta}} (T_{20})_{s's}.
\label{eq:me_spinor}
\end{align}
By inserting the result~\eqref{eq:me_spinor} into Eq.~\eqref{eq:new_para}, we arrive at
\begin{align}
\mathcal{M}_{p\Delta^{+}}[\Gamma_{q}] |_{t,\xi\to0} = \frac{4}{3} \frac{M_{N}}{M_{\Delta}} (T_{20})_{s's} H_{X}(x,0,0).
\label{eq:result}
\end{align}
Now, comparing Eq.~\eqref{eq:result} with Eq.~\eqref{eq:4}, we find the relation between the GPD and the quark distribution:
\begin{align}
H^{q}_{X}(x,0,0) &\equiv f^{q}_{Q}(x) = \frac{3}{2} \frac{M_{\Delta}}{M_{N}} f^{q}_{p\Delta^{+}}(x),
\label{eq:result2}
\end{align}
where the forward limit of the GPD $H_{X}$ has been identified as the ``quadrupole” PDF $f_{Q}$~\cite{Kim:2025ilc}. 

Furthermore, using Eqs.~\eqref{eq:smoment} and~\eqref{eq:result2}, we relate the second $x$-moment of the quark distribution to the transition EMT form factor:
\begin{align}
F^{q}_{4}(0) &= \frac{3}{4} \frac{M_{\Delta}}{M_{N}} \int^{1}_{-1} dx \, x f^{q}_{p\Delta^{+}}(x).
\label{eq:result3}
\end{align}

Lastly, using relations~\eqref{eq:result2} and~\eqref{eq:result3} and the values listed in Tab.~\ref{tab:1}, we predict the $x$ dependence of the GPD $H_{X}$ and the transition EMT form factor at $t=0$ as follows:
\begin{align}
H^{u-d}_{X}(x,0,0) &=f^{u-d}_{Q}(x)= \frac{3}{2} f^{u-d}_{p\Delta^{+}}(x), \quad \text{(see Fig.~\ref{fig:6})} \nonumber \\[1ex]
F^{u-d}_{4}(0) &= \frac{3}{4} \int^{1}_{-1} dx \, x f^{u-d}_{p\Delta^{+}}(x)\approx -0.003,
\label{eq:result4}
\end{align}
where the use of the equal mass limit $M{_N} = M_{\Delta}$ is justified by the large-$N_{c}$ kinematics. In particular, the overall strength and shape of the $x$-dependence of the unpolarized quark distribution are comparable to the prediction from Ref.~\cite{Kim:2025ilc}.

Note that the flavor singlet/octet components of the GPD $H^{q}_{X}(x,0,0)$ and the transition form factor $F^{q}_{4}(0)$ vanish. In Ref.~\cite{Alharazin:2023zzc}, isospin-symmetry breaking effects causing non-zero isoscalar EMT form factors were studied, and the smallness of these form factors was also observed in Ref.~\cite{Ozdem:2022zig}.

\section{Summary and conclusions \label{sec:8}}

In this work, we have studied the unpolarized quark distribution—also referred to as the tensor-polarized parton density—in the $N \to \Delta$ transition, using the large-$N_c$ light-cone wave function derived from the mean-field picture. We first parameterized the forward matrix element of the non-local vector operator between nucleon and $\Delta$ baryon states in terms of the parton density. Since this matrix element exhibits the tensor-polarized (quadrupole) spin structure of the baryon, the corresponding parton density is identified as the tensor-polarized parton density. Drawing an analogy to the unpolarized quark distribution in the nucleon, we have also referred to the tensor-polarized parton density in the $N \to \Delta$ transition as an unpolarized quark distribution.

To estimate this quark distribution, we constructed the baryon wave function in terms of the $3Q$, $5Q$, $7Q$, ... Fock components in the baryon rest frame using the mean-field approach. Since this wave function preserves three-dimensional rotational symmetry, the standard dynamical spin-flavor symmetry in the large-$N_c$ limit is naturally realized within this framework. However, the physical baryon wave function at rest simultaneously includes Fock components associated with both the baryon and the QCD vacuum, making it impossible to unambiguously distinguish between them. To resolve this, we employed the Lorentz invariance of the effective action and the covariance of the mean-field solution to derive the baryon wave function in the infinite momentum frame, with the spin-flavor symmetry appropriately adapted. The resulting wave function is the large-$N_c$ light-cone wave function.

Using the overlap representation of the large-$N_c$ light-cone wave function up to the $5Q$ Fock components, we have computed both diagonal and non-diagonal matrix elements of the non-local vector operator in the forward limit. We first confirmed that the general sum rules for quark distributions are well satisfied for the nucleon, the $\Delta$ baryon, and the $N \to \Delta$ transition. Importantly, for the nucleon and the $\Delta$ baryon, the leading contributions to the unpolarized quark distributions arise from the $3Q$ Fock components, with sea quark ($5Q$) contributions appearing as small corrections. In contrast, for the $p \to \Delta^{+}$ transition, the leading contribution to the quark distribution originates from the overlap of the $5Q$ Fock components. This indicates that the $p \to \Delta^{+}$ quark distribution serves as a sensitive probe of the $5Q$ structure of the baryon wave function, providing insight into the underlying chiral dynamics. Furthermore, the isospin difference between the nucleon and the $\Delta$ baryon selects the isovector component of the quark distribution, which is insensitive to gluon contributions and scale evolution.

Numerically, we have estimated the separate contributions from valence quarks, sea quarks, and antiquarks to the $p \to \Delta^{+}$ unpolarized quark distribution. These individual contributions are shown in Fig.~\ref{fig:6}, and the corresponding values of their first and second $x$-moments are listed in Tab.~\ref{tab:1}. We find that both the valence and sea quark contributions are generally small, consistent with the standard large-$N_c$ prediction~\cite{Goeke:2001tz, Kim:2024hhd}.

We then aimed to relate this quantity to known observables and provide their physical interpretations. Very recently, the $N \to \Delta$ transition GPD $H_{X}$ was introduced in Ref.~\cite{Kim:2024hhd} to complete the parameterization set of the $N \to \Delta$ transition matrix element. The first and second $x$-moments of this GPD $H_{X}$ correspond to the vanishing vector form factor and the transition energy-momentum tensor form factor $F_{4}$~\cite{Kim:2022bwn}, respectively. By taking the forward limit $H_{X}(x,0,0)$, we related it to the $p \to \Delta^{+}$ unpolarized quark distribution. Furthermore, using the $x$-moment relations between the GPD and the form factors, we connected the quark distribution to the transition energy-momentum tensor form factor $F_{4}(0)$ at zero momentum transfer.

The present mean-field picture can also be applied to study the matrix element of the non-local axial vector operator. Similar to $H_{X}$, the necessity of a new GPD, $C_{X}$, in the axial-vector sector was discussed in Ref.~\cite{Kim:2024hhd}. In this work, we found that the values of $H_{X}$ are small, confirming the standard large-$N_c$ prediction. In contrast, $C_{X}$ is not suppressed in the $1/N_c$ expansion, indicating the need for a more comprehensive analysis. Therefore, studying the $x$-dependence of the quark distribution $C_{X}$ within the present framework is an important task for future work.

\appendix
\section{Baryon rotational wave functions \label{appendix:a}}

The wave functions for the baryon octet are represented as $P_f^g$ with quark and antiquark indices, whereas those for the baryon decuplet $D_{f_1f_2f_3}$ have three quark indices. This means that the Wigner $D$ functions for the baryon octet and decuplet can be expressed as follows:
\begin{align}
&[D^{(8,1/2)*}(R)]^{g}_{f,k} \sim \epsilon_{kl} R^{\dagger l}_{f} R^{g}_{3}, \\[1ex]
&[D^{(10,3/2)*}(R)]^{g}_{\{f_1f_2f_3\},\{k_1k_2k_3\}} \nonumber \\[1ex]
&\sim \epsilon_{k'_{1}k_{1}}\epsilon_{k'_{2}k_{2}}\epsilon_{k'_{3}k_{3}} R^{\dagger k'_1}_{f_1}R^{\dagger k'_2}_{f_2}R^{\dagger k'_3}_{f_3} |_{\{f_1f_2f_3\}},
\end{align}
where $\epsilon_{kl}$ stands for the antisymmetric tensor, normalized to $\epsilon_{12}=1$. The $k=1,2 = \uparrow, \downarrow$ indices represent the spin of the baryon octet. The wave functions for the baryon decuplet are fully symmetrized in the flavor $\{f_1f_2f_3\}$ and spin $\{k_1k_2k_3\}$ indices. The flavor part of the baryon octet is explicitly written as~\cite{Oh:2004gz}:
\begin{align}
&P^{3}_{1} = p, \ \ \ P^{3}_{2} = n, 
\end{align}
and that of the decuplet $D_{\{f_1f_2f_3\}}$ is given as
\begin{align}
&D_{111} =\sqrt{6} \Delta^{++}, \quad D_{112} =\sqrt{2} \Delta^{+}, \cr
&D_{122} =\sqrt{2} \Delta^{0}, \quad D_{222} =\sqrt{6} \Delta^{-}, 
\end{align}
The prefactors of the wave functions are determined by normalizing the rotational wave functions
\begin{align}
\int dR \, B^{*}_{\mathrm{spin}}(R)B^{\mathrm{spin}}(R)=1.
\end{align}
For example, the nucleon state is explicitly expressed as
\begin{align}
  p^{*}_{k}(R)= \sqrt{8} \epsilon_{kl} R^{\dagger l}_{1}R^{3}_{3}, \ \ \
  n^{*}_{k}(R)= \sqrt{8} \epsilon_{kl} R^{\dagger l}_{2}R^{3}_{3},
\end{align}
and $\Delta$ baryons with spin-3/2 ($\uparrow \uparrow \uparrow$) and spin-1/2~($\uparrow$) are written as
\begin{align}
\Delta^{++*}_{\uparrow\uparrow\uparrow}(R)&= \sqrt{10} R^{\dagger2}_{1}R^{\dagger2}_{1}R^{\dagger2}_{1}, \nonumber \\[1ex]
\Delta^{+*}_{\uparrow\uparrow\uparrow}(R)&= \sqrt{30} R^{\dagger2}_{1}R^{\dagger2}_{1}R^{\dagger2}_{2}, \nonumber \\[1ex]
\Delta^{++*}_{\uparrow}(R)&= \sqrt{30} R^{\dagger2}_{1}R^{\dagger2}_{1}R^{\dagger1}_{1}, \nonumber \\[1ex]
\Delta^{+*}_{\uparrow}(R)&= \sqrt{10} (R^{\dagger2}_{1}R^{\dagger2}_{1}R^{\dagger1}_{2} + 2R^{\dagger2}_{2}R^{\dagger2}_{1}R^{\dagger1}_{1}).
\end{align}
For different spin projections between the initial and final baryon wave functions, the integral is zero. The rotational wave functions belonging to different baryons are also orthogonal.

\section{Quark distributions for $\Delta^{+}$ baryon \label{appendix:b}}
We summarize the results of the quark distributions for the $\Delta^{+}$ baryon as follows [here, we suppress the dependence on $x$, i.e., $\mathcal{K}(x) \equiv \mathcal{K}$]:
\begin{itemize}
\item Active $3Q$ and inactive $Q\bar{Q}$ pair
\begin{align}
O^{u_{\mathrm{val}},(5)}_{\Delta^{+}}&=\frac{3}{80} \left[ 135\mathcal{K}_{\sigma \sigma} + 73\mathcal{K}_{\pi \pi}   + 30\mathcal{K}_{33}  \right], \cr
O^{d_{\mathrm{val}},(5)}_{\Delta^{+}}&=\frac{3}{20} \left[ 17\mathcal{K}_{\sigma \sigma} + 11\mathcal{K}_{\pi \pi}   + 6\mathcal{K}_{33} \right], \cr
O^{s_{\mathrm{val}},(5)}_{\Delta^{+}}&=\frac{3}{80} \left[ \mathcal{K}_{\sigma \sigma} + 15\mathcal{K}_{\pi \pi}  +  18\mathcal{K}_{33}  \right].
\end{align} 
\item Inactive $3Q$ and active quark in the $Q\bar{Q}$ pair
\begin{align}
O^{u_{\mathrm{s}},(5)}_{\Delta^{+}}&=\frac{3}{80} \left[ 29\mathcal{K}^{+}_{\sigma \sigma} + 27\mathcal{K}^{+}_{\pi \pi} + 18\mathcal{K}^{+}_{33}  \right], \cr
O^{d_{\mathrm{s}},(5)}_{\Delta^{+}}&=\frac{1}{20} \left[ 17\mathcal{K}^{+}_{\sigma \sigma} + 11\mathcal{K}^{+}_{\pi \pi}  + 6\mathcal{K}^{+}_{33}  \right], \cr
O^{s_{\mathrm{s}},(5)}_{\Delta^{+}}&=\frac{1}{80} \left[ 49\mathcal{K}^{+}_{\sigma \sigma} + 7\mathcal{K}^{+}_{\pi \pi}  -6\mathcal{K}^{+}_{33}  \right].
\end{align} 
\item Inactive $3Q$ and active antiquark in the $Q\bar{Q}$ pair
\begin{align}
O^{\bar{u},(5)}_{\Delta^{+}}&=\frac{3}{20} \left[ 7\mathcal{K}^{-}_{\sigma \sigma} + 3\mathcal{K}^{-}_{\pi \pi}  \right], \cr
O^{\bar{d},(5)}_{\Delta^{+}}&=\frac{1}{20} \left[ 17\mathcal{K}^{-}_{\sigma \sigma} + 11\mathcal{K}^{-}_{\pi \pi}  + 6\mathcal{K}^{-}_{33}  \right], \cr
O^{\bar{s},(5)}_{\Delta^{+}}&=\frac{1}{20} \left[ 13\mathcal{K}^{-}_{\sigma \sigma} + 13\mathcal{K}^{-}_{\pi \pi}  + 12\mathcal{K}^{-}_{33}  \right].
\end{align} 
\end{itemize}
Similar to the proton, we observe that the baryon number and momentum sum rules for the $\Delta^{+}$ baryon are respectively satisfied:
\begin{align}
&\int^{1}_{-1} dx \, f^{u}_{\Delta^{+}}(x) = 2, \quad \int^{1}_{-1} dx \, f^{d}_{\Delta^{+}}(x) = 1, \cr
&\int^{1}_{-1} dx \, f^{s}_{\Delta^{+}}(x) = 0,
\end{align}
and
\begin{align}
&\int^{1}_{-1} dx \, x f^{u+d+s}_{\Delta^{+}}(x) = 1.
\end{align}
%
\section*{Acknowledgments}
The author is indebted to C. Weiss for valuable discussions and for his advice on the presentation and content of the manuscript. This material is based upon work supported by the U.S.~Department of Energy, Office of Science, Office of Nuclear Physics under contract DE-AC05-06OR23177. 
\bibliography{NDeltaPDFs}
\end{document}